\documentclass[a4paper,twoside,british,useAMS,usenatbib]{mn2e}
\usepackage{lmodern}

\usepackage[T1]{fontenc}
\usepackage[latin9]{inputenc}
\usepackage{babel}
\usepackage{graphicx}
\usepackage[authoryear]{natbib}
\usepackage[unicode=true,
 bookmarks=true,bookmarksnumbered=false,bookmarksopen=false,
 breaklinks=false,pdfborder={0 0 0},backref=false,colorlinks=false]
 {hyperref}
\hypersetup{
 pdfauthor={Lee S. Kelvin},
 pdfsubject={Extragalactic Astronomy}}

\makeatletter

\pdfpageheight\paperheight
\pdfpagewidth\paperwidth

\newcommand{\noun}[1]{\textsc{#1}}

\renewcommand\[{\begin{equation}}
\renewcommand\]{\end{equation}}
\usepackage{aas_macros}

\makeatother

\begin{document}
\title[GAMA: Stellar mass functions by Hubble type]{Galaxy And Mass Assembly (GAMA): Stellar mass functions by Hubble type}

\author[L.~S.~Kelvin et al.]{
\parbox{\textwidth}{
\raggedright
Lee~S.~Kelvin,$^{1,2,3}$
Simon~P.~Driver,$^{2,3}$
Aaron~S.~G.~Robotham,$^{3}$
Edward~N.~Taylor,$^{4}$
Alister~W.~Graham,$^{5}$
Mehmet~Alpaslan,$^{2}$
Ivan~Baldry,$^{6}$
Steven~P.~Bamford,$^{7}$
Amanda~E.~Bauer,$^{8}$
Joss~Bland-Hawthorn,$^{9}$
Michael~J.~I.~Brown,$^{10}$
Matthew~Colless,$^{11}$
Christopher~J.~Conselice,$^{7}$
Benne~W.~Holwerda,$^{12}$
Andrew~M.~Hopkins,$^{8}$
Maritza~A.~Lara-López,$^{8}$
Jochen~Liske,$^{13}$
Ángel~R.~López-Sánchez,$^{8,14}$
Jon~Loveday,$^{15}$
Peder~Norberg,$^{16}$
Steven~Phillipps,$^{17}$
Cristina~C.~Popescu,$^{18}$
Matthew~Prescott,$^{19}$
Anne~E.~Sansom,$^{18}$
and Richard~J.~Tuffs$^{20}$
}\vspace{0.4cm}\\
\parbox{\textwidth}{
$^{1}$Institut f\"{u}r Astro- und Teilchenphysik, Universit\"{a}t Innsbruck, Technikerstra{\ss}e 25, 6020 Innsbruck, Austria\\
$^{2}$School of Physics and Astronomy, University of St Andrews, North Haugh, St Andrews, Fife, KY16 9SS, UK\\
$^{3}$International Centre for Radio Astronomy Research, 7 Fairway, The University of Western Australia, Crawley, Perth, Western Australia 6009, Australia\\
$^{4}$School of Physics, the University of Melbourne, Parkville, VIC 3010, Australia\\
$^{5}$Centre for Astrophysics and Supercomputing, Swinburne University of Technology, Hawthorn, Victoria 3122, Australia\\
$^{6}$Astrophysics Research Institute, Liverpool John Moores University, Twelve Quays House, Egerton Wharf, Birkenhead CH41 1LD, UK\\
$^{7}$School of Physics and Astronomy, The University of Nottingham, University Park, Nottingham, NG7 2RD, UK\\
$^{8}$Australian Astronomical Observatory, PO Box 915, North Ryde, NSW 1670, Australia\\
$^{9}$Sydney Institute for Astronomy, School of Physics A28, University of Sydney, NSW 2006, Australia\\
$^{10}$School of Physics, Monash University, Clayton, VIC 3800, Australia\\
$^{11}$Research School of Astronomy and Astrophysics, The Australian National University, Canberra, ACT 2611, Australia\\
$^{12}$European Space Agency, ESTEC, Keplerlaan 1, NL-2200 AG Noordwijk, The Netherlands\\
$^{13}$European Southern Observatory, Karl-Schwarzschild-Str. 2, 85748 Garching, Germany\\
$^{14}$Department of Physics and Astronomy, Macquarie University, NSW 2109, Australia\\
$^{15}$Astronomy Centre, University of Sussex, Falmer, Brighton BN1 9QH, UK\\
$^{16}$Institute for Computational Cosmology, Department of Physics, Durham University, South Road, Durham DH1 3LE, UK\\
$^{17}$Astrophysics Group, H.H. Wills Physics Laboratory, University of Bristol, Tyndall Avenue, Bristol BS8 1TL, UK\\
$^{18}$Jeremiah Horrocks Institute, School of Computing, Engineering and Physical Sciences, University of Central Lancashire, Preston PR1 2HE, UK\\
$^{19}$Astrophysics Group, Department of Physics, University of the Western Cape, 7535 Bellville, Cape Town, South Africa\\
$^{20}$Max Planck Institut f\"{u}r Kernphysik, Saupfercheckweg 1, D-69117 Heidelberg, Germany\\
}
\vspace{-0.75cm}
}

\date{Accepted for publication in MNRAS}

\pagerange{\pageref{firstpage}--\pageref{lastpage}} \pubyear{2013}

\maketitle

\label{firstpage}
\begin{abstract}
We present an estimate of the galaxy stellar mass function and its
division by morphological type in the local ($0.025<z<0.06$) Universe.
Adopting robust morphological classifications as previously presented
(\citeauthor{Kelvin2014a}) for a sample of $3,727$ galaxies taken
from the Galaxy And Mass Assembly survey, we define a local volume
and stellar mass limited sub-sample of $2,711$ galaxies to a lower
stellar mass limit of $\mathcal{M}=10^{9.0}\mathcal{M}_{\odot}$.
We confirm that the galaxy stellar mass function is well described
by a double Schechter function given by $\mathcal{M}^{*}=10^{10.64}\mathcal{M}_{\odot}$,
$\alpha_{1}=-0.43$, $\phi_{1}^{*}=4.18\;\mathrm{dex}^{-1}\mathrm{Mpc}^{-3}$,
$\alpha_{2}=-1.50$ and $\phi_{2}^{*}=0.74\;\mathrm{dex}^{-1}\mathrm{Mpc}^{-3}$.
The constituent morphological-type stellar mass functions are well
sampled above our lower stellar mass limit, excepting the faint little
blue spheroid population of galaxies. We find approximately $71{}_{-4}^{+3}\%$
of the stellar mass in the local Universe is found within spheroid
dominated galaxies; ellipticals and S0-Sas. The remaining $29{}_{-3}^{+4}\%$
falls predominantly within late type disk dominated systems, Sab-Scds
and Sd-Irrs. Adopting reasonable bulge-to-total ratios implies that
approximately half the stellar mass today resides in spheroidal structures,
and half in disk structures. Within this local sample, we find approximate
stellar mass proportions for E : S0-Sa : Sab-Scd : Sd-Irr of $34$
: $37$ : $24$ : $5$.
\end{abstract}
\begin{keywords}
Galaxies -- galaxies: elliptical and lenticular, cD -- galaxies: spiral -- galaxies: luminosity function, mass function -- galaxies: fundamental parameters
\vspace{-0.5cm}
\end{keywords}

\section{Introduction}

\label{sec:intro}Amongst the veritable cornucopia of observed and
derived galaxy parameters, the total stellar mass of a system is arguably
one of the most fundamental, perhaps in conjunction with the shape
of the galaxy light profile as parameterised by, e.g., the \citet{Sersic1963}
function. One common viewpoint has it that galaxies form via a series
of monolithic collapse and/or hierarchical merging events, whereafter
evolution continues to occur via additional merging events and stochastic
gas accretion phases (e.g., \citealp{Navarro1991,White1991,Cook2009,LHuillier2012,Khochfar2006a,DeLucia2007,Pichon2011,Wyse1997,vanDokkum2010a,Khochfar2006b,Kormendy2012,Keres2005,Cook2010a,Debattista2006}).
Each stage during this galactic ageing process has an observational
impact upon the instantaneous state of a galaxy, e.g.; colour \citep{Baldry2004,Baldry2008a},
star formation rate \citep{Behroozi2013,Moustakas2013}, in addition
to leaving a longer term imprint on the nature of the galaxy, e.g.;
metallicity \citep{DeLucia2004,Driver2013,Lara-Lopez2013}, structure
\citep{Cooper2012,Shankar2013,Szomoru2013,Robotham2013}. In many
ways this latter parameter, galaxy structure, promises to be the most
profound, as rearranging the orbital properties of billions of stars
is not a whimsical thing.

Several well known relations between stellar mass and additional complementary
galaxy parameters are known to exist, including total size (\citealp{Graham2006b,Patel2013}),
velocity dispersion (\citealp{Davies1983a,Davies1983b,Shen2003,Matkovic2005}),
concentration indices and light profile shapes (\citealp{Caon1993,Young1994,Kauffmann2003b,Blanton2005a,Kelvin2012}),
environment (\citealp{Kauffmann2004,Baldry2006}), metallicity (\citealp{Tremonti2004}),
metallicity and star formation rate in a 3-dimensional plane (\citealp{Lara-Lopez2010})
and colour (\citealp{Conselice2006b}). This latter study highlights
the importance of stellar mass above other observed parameters, such
as star formation rate and merger activity, in describing the maximal
variance across the galaxy population. Numerous recent studies explore
the division of the local stellar mass budget by, e.g., colour (\citealp{Baldry2012,Peng2012};
Taylor et al., 2014, submitted), star formation rate (\citealp{Pozzetti2010}),
environment (\citealp{Bolzonella2010}) and coarse morphology (\citealp{Bundy2010}).
Here we study the relation between stellar mass and morphology, specifically;
how the local galaxy stellar mass function (GSMF) is built from different
morphological types. A standard cosmology of ($H_{0}$, $\Omega_{m}$,
$\Omega_{\Lambda}$)$=$($70$ km s$^{-1}$ Mpc$^{-1}$, $0.3$, $0.7$)
is assumed throughout this paper.

\section{Data}

\label{sec:data}Our data is taken from the Galaxy And Mass Assembly
survey \citep[GAMA:][]{Driver2009,Driver2011} phase $1$ (GAMA I).
GAMA is a combined spectroscopic and multi-wavelength imaging survey
designed to study spatial structure in the nearby ($z<0.25$) Universe
on scales of $1$ kpc to $1$ Mpc. The GAMA regions lie within areas
of sky previously surveyed by both SDSS (\citealt{York2000,Abazajian2009})
as part of its Main Survey, and UKIRT as part of the UKIDSS Large
Area Survey (UKIDSS-LAS; \citealp{Lawrence2007}).

Using the latest version (version $16$) of the GAMA I tiling catalogue%
\footnote{All data release 2 GAMA catalogues are available through the GAMA
database, available online at http://www.gama-survey.org/dr2/ .%
} (\emph{TilingCatv16}, see \citealp{Baldry2010}), we adopt a local,
volume and luminosity limited sample of $3,727$ galaxy-like ($\mathrm{SURVEY\_CLASS}\geq2$)
objects, GAMAnear, previously defined in \citet{Kelvin2014a}. In
brief, this sample is defined thus: 
\begin{itemize}
\item a local flow-corrected spectroscopic redshift $z$ of $0.025<z<0.06$
with an associated GAMA redshift quality flag of $nQ>2$ (i.e., good
for science),
\item a Milky Way dust extinction corrected apparent $r$ band SDSS (DR7)
Petrosian magnitude of $r<19.4$ mag,
\item an absolute Sérsic magnitude truncated at $10$ multiples of the half-light
radius in the $r$-band of $M_{r}<-17.4$ mag.
\end{itemize}
Local flow-corrected spectroscopic redshifts are taken from the GAMA
I \emph{DistancesFramesv07} catalogue \citep{Baldry2012}. For this
sample, we adopt an upper redshift limit of $z=0.06$. This limit
is chosen such that the majority of bulges (the limiting structural
component) should remain resolvable%
\footnote{Assuming of course a sufficiently high B/T ratio which allows for
the detection of bulge flux above the host disk flux.%
}. To calculate this limit, typical bulge half-light radii for galaxies
in the local Universe are estimated based on prior bulge-disk decompositions
presented in \citealp{Allen2006} ($\sim1.93$ kpc) and \citealp{Simard2011}
($\sim3.02$ kpc, see \citealp{Kelvin2014a} for a complete discussion).
Our upper redshift limit is determined using these data to estimate
the maximal redshift out to which a bulge would remain larger than
the average seeing found in SDSS imaging ($\sim1.1''$, see \citealp{Kelvin2012}).
At $z=0.06$, $1.1''$ corresponds to a physical size of $1.28$ kpc.
Therefore, bulge half-light diameters are at least three times the
median $r$ band seeing at $z=0.06$. A lower limit of $z=0.025$
is also adopted to avoid low galaxy number densities below this redshift
and such that redshifts are not dominated by peculiar velocities.
(see \citealt{Kelvin2014a} for further details). Our redshift limits
give this sample a volume of $2.1\times10^{5}$ Mpc$^{3}$. Matching
the GAMAnear sample to the GAMA galaxy group catalogue (G3C; \citealp{Robotham2011b}),
we find that just under half ($1797$, $\sim48\%$) of our galaxies
lie in identified groups, with a median halo mass of $\mathcal{M}_{H}\sim10^{12.9}\mathcal{M}_{\odot}$.
Of these galaxies, $672$ ($\sim18\%$) lie in groups with a richness
greater than $5$, with a median halo mass of $\mathcal{M}_{H}\sim10^{13.5}\mathcal{M}_{\odot}$.
Owing to this, our sample should be considered predominantly field
dominated, extending into the intermediate-mass group regime.

Our SDSS DR7 (\citealt{York2000,Abazajian2009}) apparent Petrosian
magnitude limit of $r=19.4$ is chosen to correspond to the main GAMA
I spectroscopic target selection limit \citep{Driver2009,Baldry2010},
ensuring completeness across all three equatorial GAMA regions%
\footnote{Whilst the central 12h equatorial GAMA field (G12) reaches a greater
limiting depth of $r=19.8$, we choose not to consider this extra
data here to maintain a consistent depth of $r=19.4$ across all three
primary GAMA fields.%
}. Sérsic magnitudes are robustly derived using the galaxy 2D light-profile
modelling package SIGMA \citep{Kelvin2010,Kelvin2012}. Information
on their derivation, and a further discussion of our choice to truncate
these extrapolated light-profile fits to $10$ multiples of the half
light radius may be found in \citet{Kelvin2012}. Our absolute Sérsic
magnitude $r$-band limit of $M_{r}<-17.4$ mag is chosen to avoid
the effects of Malmquist bias out to our upper redshift limit of $z=0.06$.
A further discussion of this limit can be found in \citet{Kelvin2014a}.

The GAMAnear dataset is visually morphologically classified in \citet{Kelvin2014a}
by three independent observers into their appropriate Hubble type,
namely; elliptical (E), lenticular/early-type spiral (S0-Sa, barred
and unbarred), intermediate/late-type spiral (Sab-Scd, barred and
unbarred), disk-dominated spiral or irregular (Sd-Irr), star (see
below) and little blue spheroid (LBS). Classifications are assigned
on a majority agreement basis; at least two of the three observers
must agree on the classification. In the result of a three-way tie
(only occurring for $56$ galaxies, or $1.5\%$ of the total sample),
preference is given to the senior observer.

As previously noted in \citet{Kelvin2014a}, the LBS type galaxy is
typically compact, spheroidal and blue, hence their designation as
`Little Blue Spheroids'. The median colour of LBS galaxies within
our GAMAnear sample is $g-i\sim0.6$ with a median Sérsic index of
$n_{r}\sim1.9$ in the $r$ band ($n_{K}\sim1.6$ in the $K$ band)
and a median physical size of $r_{e}\sim1.1$ kpc in the $r$ band
($r_{e}\sim0.9$ kpc in the $K$ band). Because of their physical
similarities with both spheroids and disks, it is not immediately
apparent which structural class these objects should be associated
with. For a further discussion of our LBS class, we refer the reader
to \citet{Kelvin2014a}, and we note that a dedicated study is currently
in progress in order to better understand our LBS population (Moffett
et al., in prep.).

We acknowledge the apparent difficulty in visually dividing galaxies
along the elliptical/lenticular interface, as highlighted in the recent
literature, e.g., \citet{Bamford2009,Emsellem2011,Cappellari2011a,Cappellari2013}.
A face on lenticular galaxy may appear, even to the expert classifier,
as a smooth 1-component system, and therefore be mis-classified as
an elliptical galaxy. As a consequence of this, the S0-Sa class will
suffer from a shortfall in the correct number of S0 type galaxies.
Nevertheless, in keeping with the classification methodology of our
original study (\citealp{Kelvin2014a}), here we opt to maintain this
division between elliptical and lenticular type galaxies. 

The latter `star' type refers to incorrectly targeted foreground stars
in front of a background galaxy (to which the associated redshift
belongs) or segments of a large diffuse galaxy, and therefore this
population shall be discarded from subsequent morphological analyses.
Owing to low number statistics for our barred systems, the barred
populations have been merged into their sibling unbarred classes.
Any subsequent discussion of the barred populations alone are provided
for completeness sake, in keeping with the classification criteria
established in \citet{Kelvin2014a}, but this information is not used
for scientific analyses. For further details on our morphological
dataset and base sample selection criteria, see \citet{Kelvin2014a}.

\subsection{Stellar Masses}

\label{sub:stellarmasses}Stellar masses used in this study are taken
from version $8$ of the GAMA I stellar mass catalogue (\emph{StellarMassesv08};
\citealp{Taylor2011}). In summary, a series of \citet{Bruzual2003}
composite stellar population spectral models are created, adopting
a \citet{Chabrier2003} Initial Mass Function and using a \citet{Calzetti2000}
dust attenuation law. A stellar population library is constructed
under the assumptions of a single metallicity and a continuous exponentially
declining star formation history for each stellar population. Dust
is modelled as a single uniform screen. These spectra are subsequently
rescaled by some normalisation factor in order that the synthetic
spectral flux matches that as defined by a series of Kron-like (AUTO)
apertures as detailed in \citet{Hill2011}. The value of the normalisation
factor determines the AUTO aperture defined stellar mass for that
particular system, $\mathcal{M}_{*,\mathrm{AUTO}}$. 

We apply a secondary Sérsic flux correction to the AUTO defined stellar
masses as recommended in \citet{Taylor2011}. As shown in \citet{Graham2005a},
both Petrosian and Kron-like photometry have the potential to miss
flux in the wings of large, extended systems (particularly those with
high Sérsic indices). Sérsic photometry is ideally suited to correct
for this effect, and so we choose to apply it to these data. Our final
stellar mass estimates, $\log\left(\frac{\mathcal{M}_{*}}{\mathcal{M}_{\odot}}\right)$,
are given using the equation
\[
\log\left(\frac{\mathcal{M}_{*}}{\mathcal{M}_{\odot}}\right)=\log\left(\frac{\mathcal{M}_{*\mathrm{,AUTO}}}{\mathcal{M}_{\odot}}\right)+\log\left(\frac{L_{\mathrm{S\acute{e}rsic}}}{L_{\mathrm{AUTO}}}\right)
\]
where $L_{AUTO}$ and $L_{S\acute{e}rsic}$ are the (linear) r-band
AUTO aperture flux and the total r-band flux inferred from fitting
a Sérsic profile truncated at 10 multiples of the half-light radius
(as given in \citealp{Kelvin2012}), respectively. The scale factor
$L_{\mathrm{S\acute{e}rsic}}/L_{\mathrm{AUTO}}$ describes the additional
flux given by our single Sérsic model fits relative to the standard
GAMA AUTO photometry. For each morphological type we find the following
median Sérsic--AUTO flux scale factors; LBS=$1.01$, E=$1.03$, S0-Sa=$1.05$,
Sab-Scd=$1.01$, Sd-Irr=$1.00$. Note that our resultant stellar mass
estimates refer to the stellar mass implied via the visible flux from
the living stellar population within a galaxy, and not the total living
plus faint/dark remnant (i.e., white dwarf, neutron star, black hole,
etc.) populations \citep{Shimizu2013}.

As expanded upon in \citet{Baldry2012}, the GAMAnear sample will
suffer from surface brightness limitations at the faint/low-mass end
of our sample owing to photometric incompleteness. Figure $11$ of
\citet{Baldry2012} shows the relation between surface brightness
and stellar mass for a subset of the GAMA dataset across a similar
redshift range. Clearly, the impact of surface brightness incompleteness
becomes increasingly severe in the mass range $\log\left(\mathcal{M}_{*}/\mathcal{M}_{\odot}\right)=8.0-9.0$.
To mitigate the effects of incompleteness, we adopt the extreme of
this range and that recommended in \citet{Baldry2012}, $\log\left(\mathcal{M}_{*}/\mathcal{M}_{\odot}\right)=9.0$,
as an additional stellar mass limit to our sample. This reduces our
GAMAnear sample from $3,727$ galaxies to $2,711$ ($73$\% of the
GAMAnear dataset). 

The well-established relation between colour and mass-to-light ratio
(e.g., Figure $12$, \citealp{Taylor2011}) implies that galaxies
with a higher mass-to-light ratio tend towards being redder in colour.
Therefore, for a given luminosity, redder galaxies appear more massive
than their bluer counterparts. Consequently, for galaxies in our volume
and $r$ band magnitude limited GAMAnear sample, at a given stellar
mass, one is able to see bluer galaxies out to a higher redshift than
red systems (e.g., Figure A1, \citealp{vandenBosch2008}). Alternatively,
at a given redshift, the stellar mass completeness limit is higher
for red galaxies than for blue. In order to fully account for any
potential incompleteness bias within our remaining mass-limited sample
of $2,711$ galaxies, we also opt to weight each galaxy above our
mass limit according to $W=V_{\mathrm{tot}}/V_{\mathrm{max}}$ \citep{Schmidt1968};
the ratio of the total observed volume ($2.1\times10^{5}$ Mpc$^{3}$)
to the maximum comoving volume over which the galaxy could have been
observed within the survey limits. The corresponding $z_{\mathrm{max}}$
is the maximum redshift at which a galaxy can be seen based on its
spectral shape and survey limits ($r=19.4$ mag). We adopt $z_{\mathrm{max}}$
values as presented in \citet{Taylor2011} and available in the GAMA
StellarMassesv08 catalogue in order to calculate $V_{\mathrm{max}}$
estimates. Weights in the range $W<1$, i.e., $V_{\mathrm{tot}}<V_{\mathrm{max}}$,
are set equal to $1$. All stellar masses presented hereafter should
be assumed to have this $V_{\mathrm{max}}$ weight correction applied,
unless otherwise stated.

\begin{figure}
\includegraphics[width=1\columnwidth]{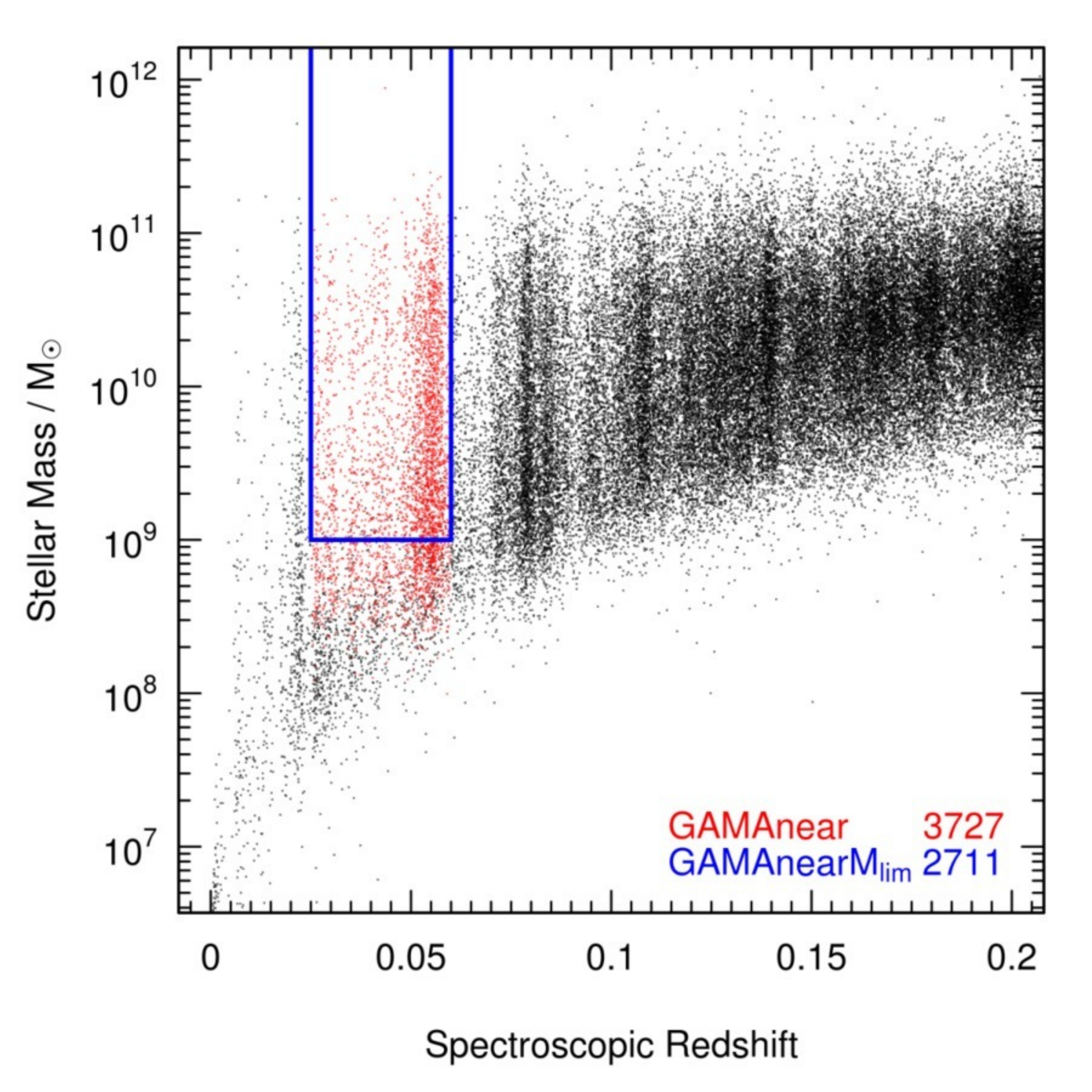}

\caption{\label{fig:massgamanear}Stellar mass as a function of redshift for
galaxies within the GAMA survey. The red data points represent our
GAMAnear sample; those galaxies that lie in the redshift range $0.025<z<0.06$
with an associated GAMA redshift quality flag of $nQ>2$ (i.e., good
for science), an extinction corrected $r$ band SDSS Petrosian magnitude
of $r<19.4$ mag and an absolute truncated Sérsic magnitude in the
$r$-band of $M_{r}<-17.4$ mag. GAMAnear galaxies that additionally
are more massive than $\log\left(\mathcal{M}_{*}/\mathcal{M}_{\odot}\right)>9.0$
(within the blue box) constitute our stellar mass limited sample,
GAMAnearM$_{\mathrm{lim}}$, in use throughout the remainder of this
paper. Stellar masses shown here are not $V_{\mathrm{max}}$ weight
corrected.}

\end{figure}

This volume and stellar mass limited sample of $2,711$ galaxies,
GAMAnear$\mathcal{M}_{\mathrm{lim}}$, constitutes our primary dataset,
and shall be used throughout the remainder of this paper. Both GAMAnear
and GAMAnear$\mathcal{M}_{\mathrm{lim}}$ are shown in redshift--stellar
mass space in Figure \ref{fig:massgamanear}.

\section{Stellar Mass and Morphology}

\label{sec:massmorphology}
\begin{figure*}
\includegraphics[width=1\textwidth]{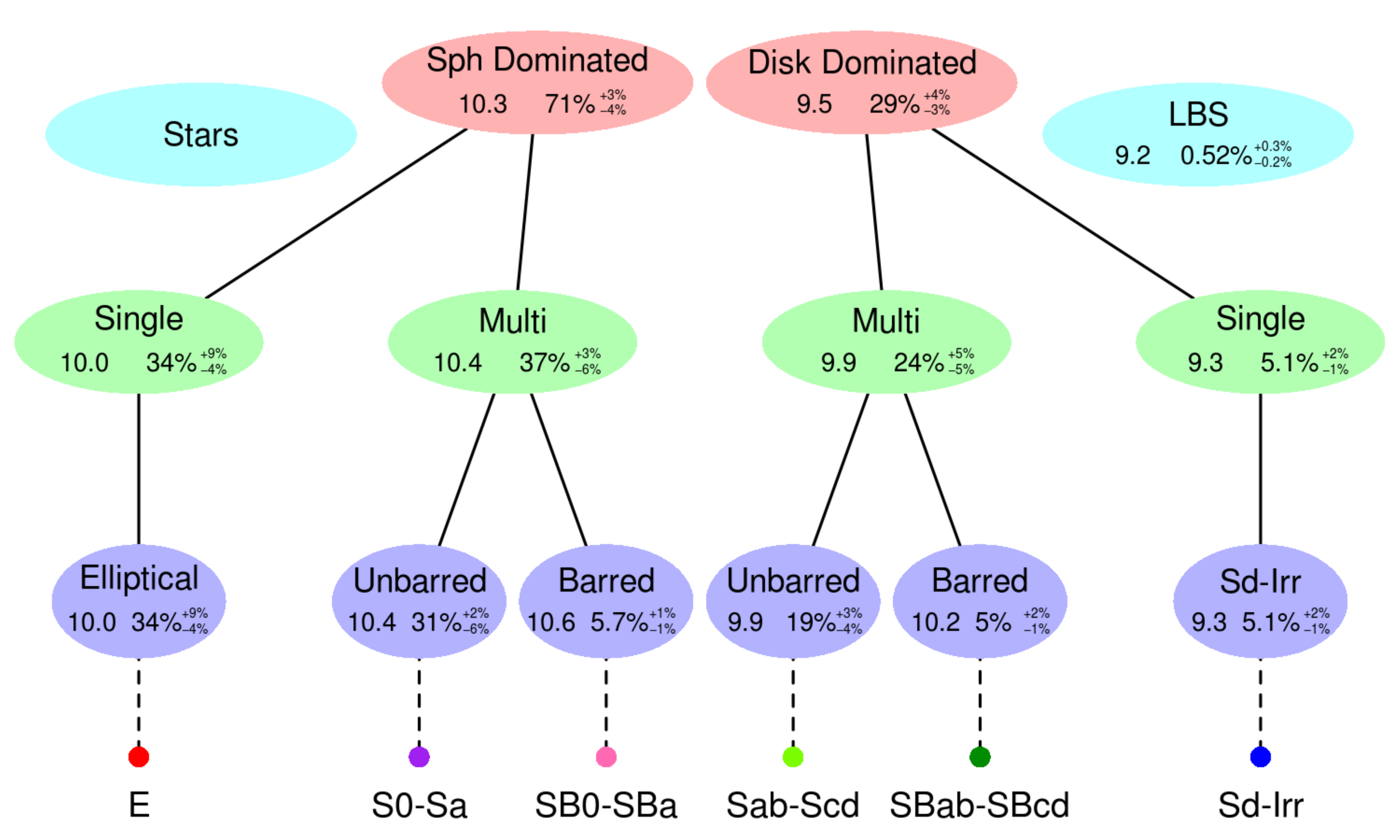}

\caption{\label{fig:masscount}The breakdown of stellar mass within our GAMAnear$\mathcal{M}_{\mathrm{lim}}$
sample by morphological type. Within each classification bubble, values
for the logarithm of the median stellar mass (left) and the percentage
of total stellar mass with its associated error (right) are shown.
Stellar masses shown here have been $V_{\mathrm{max}}$ weight corrected.}

\end{figure*}

Figure \ref{fig:masscount} shows the stellar
mass breakdown by type and morphology for the entirety of our mass
limited sample of $2,711$ galaxies. Within each classification bubble,
values for the logarithm of the median stellar mass (left) and the
percentage by stellar mass with associated error (right) are shown.
Percentage by stellar mass is calculated via a simple summation of
the $V_{\mathrm{max}}$-weighted stellar mass of each galaxy within
each population. Percentage errors represent the maximal dispersion
between the three independent classifiers, i.e., the stellar mass
for each galaxy population is rederived for each classifier and the
offset from the master classification calculated. Note that the stellar
masses for each galaxy as derived in \citet{Taylor2011} have a typical
associated intrinsic stellar mass error of $\Delta\log\left(\mathcal{M}_{*}/\mathcal{M}_{\odot}\right)\sim0.1$. 

Approximately $71{}_{-4}^{+3}$\% of the stellar mass in our GAMAnear$\mathcal{M}_{\mathrm{lim}}$
sample is currently found within spheroid dominated%
\footnote{Here, the term `spheroid dominated' does not refer to the spheroidal
component dominating the total flux of the system. As has been shown
in \citet{Graham2008b}, rarely does the spheroid component in a bulge+disk
system contribute $>50$\% of the flux for galaxies later than S0.
Rather, we define the term `spheroid dominated' to refer to the visual
impact of the spheroid on the postage stamp images presented in \citet{Kelvin2014a};
a combination of relative size, apparent surface brightness and 2D
light profile.%
} (elliptical and S0-Sa) systems, with the remaining stellar mass in
disk dominated (Sab-Scd and Sd-Irr) galaxies ($29{}_{-3}^{+4}$\%)
and little blue spheroids ($\sim0.52\%_{-0.2}^{+0.3}$). Adopting
reasonable bulge-to-total values (e.g., for an intermediate Sb spiral,
$\log\left(\mathrm{B}/\mathrm{D}\right)\sim-1$, \citealp{Graham2008b})
implies that approximately half the stellar mass today resides in
spheroidal structures%
\footnote{One expects the bulge-to-total ratio to correlate with the total stellar
mass of the system, and therefore, this value should be considered
an estimate.%
}, with the remaining half within disk-like structures, in-line with
previous studies (see \citealp{Driver2007a,Driver2007b,Gadotti2009,Tasca2011}).
Continuing further down the classification tree, we find approximate
stellar mass proportions for E : S0-Sa : Sab-Scd : Sd-Irr of $34$
: $37$ : $24$ : $5$. For comparison, table \ref{tab:massfracs}
shows the number fractions of various galaxy populations in stellar
mass ranges with progressively more massive lower bounds. We see that
no LBS type galaxies exist in the mass range $\log\left(\mathcal{M}_{*}/\mathcal{M}_{\odot}\right)>10.0$.
Spheroid dominated galaxies become more numerous than disk dominated
galaxies at $\log\left(\mathcal{M}_{*}/\mathcal{M}_{\odot}\right)>9.5$,
whilst elliptical galaxies alone dominate the galaxy population by
number at $\log\left(\mathcal{M}_{*}/\mathcal{M}_{\odot}\right)>11.0$.
Interestingly, at stellar masses less massive than $\log\left(\mathcal{M}_{*}/\mathcal{M}_{\odot}\right)=11.0$,
$\sim25$\% of the total galaxy population are consistently elliptical.

\renewcommand{\arraystretch}{1.5}
\setlength{\tabcolsep}{5pt}

\begin{table}
\begin{centering}
\begin{tabular}{ccccccc}
\hline 
Population & \multicolumn{6}{c}{Stellar Mass Range [$\log\left(\mathcal{M}_{*}/\mathcal{M}_{\odot}\right)$]} \tabularnewline
           & $>9$ & $>9.5$ & $>10$ & $>10.5$ & $>11$ & $>11.5$ \tabularnewline
\hline 
LBS & $0.04$ & $0.01$ & $0.00$ & $0.00$ & $0.00$ & $0.00$ \tabularnewline
E & $0.19$ & $0.24$ & $0.24$ & $0.31$ & $0.72$ & $1.00$ \tabularnewline
S0-Sa & $0.18$ & $0.28$ & $0.42$ & $0.48$ & $0.22$ & $0.00$ \tabularnewline
Sab-Scd & $0.28$ & $0.36$ & $0.32$ & $0.20$ & $0.06$ & $0.00$ \tabularnewline
Sd-Irr & $0.31$ & $0.11$ & $0.02$ & $0.00$ & $0.00$ & $0.00$ \tabularnewline
\hline
Sph Dom & $0.37$ & $0.51$ & $0.66$ & $0.80$ & $0.94$ & $1.00$ \tabularnewline
Disk Dom & $0.59$ & $0.47$ & $0.34$ & $0.20$ & $0.06$ & $0.00$ \tabularnewline

\hline 
\end{tabular}
\par\end{centering}

\caption{\label{tab:massfracs}The number fractions of various galaxy populations
for stellar mass ranges with progressively more massive lower bounds,
as indicated. Barred populations have been merged into their sibling
unbarred classes. Stellar masses shown here have been $V_{\mathrm{max}}$
weight corrected.}
\end{table}

\renewcommand{\arraystretch}{1}

Our elliptical stellar mass fraction of $34\%_{-4}^{+9}$ is in excellent
agreement with the value of $32\%$ found in \citet{Gadotti2009}%
\footnote{Note that whilst the sample in \citealt{Gadotti2009} spans a similar
redshift range, their lower stellar mass limit is $1$ dex higher,
$10^{10}$ $\mathcal{M}_{\odot}$, than that adopted here.%
} but significantly higher than the $\sim15$\% value found in \citet{Driver2007a}.
This presumably reflects the great difficulty in distinguishing between
genuine pressure supported ellipticals and rotationally supported
face-on lenticulars, as highlighted by the \noun{ATLAS3D} team, see
for example \citet{Emsellem2011,Krajnovic2011,Cappellari2011b,Duc2011,Khochfar2011},
also \citet{D'Onofrio1995,Graham1998}. If the \citeauthor{Driver2007a}
study is correct then the potential contamination of our elliptical
class by lenticular types may be significant. A key difference in
our classifications and that of \citeauthor{Driver2007a} is the method
of selection, with the former using eyeball morphology based on SDSS/UKIDSS
data and the latter using \noun{GIM2D} bulge-disc decompositions based
on the significantly deeper Millennium Galaxy Catalogue $B$ band
data (see \citealp{Liske2003b,Driver2005}). The \citet{Gadotti2009}
elliptical class is based on a Petrosian concentration index cut.
In \citet{Driver2006} it was reported than the E/S0 (red spheroid)
class contains $\left(35\pm2\right)$\% of the stellar mass, which
is closer to our elliptical value, and perhaps supporting the notion
that our visually classified E class potentially contains a large
fraction of lenticular contaminants. We will explore this issue in
detail using robust structural decompositions (Kelvin et al., in prep.)
based on the GALFIT galaxy fitting software \citep{Peng2002,Peng2010a}
and via ongoing SAMI and CALIFA integral field unit observations (in
progress). At present we advocate a small amount of caution in regards
to the level of potential lenticular contamination of our elliptical
sample.

\section{The Stellar Mass Function}

\label{sec:massfuncs}

\subsection{The Galaxy Stellar Mass Function}

\label{sub:schechter}One of the most fundamental measurements in
astronomy is that of the galaxy luminosity function, or its equivalent
in mass, the galaxy stellar mass function (hereafter GSMF). The GSMF
gives the effective number of galaxies per unit volume in the logarithmic
stellar mass interval $\log\mathcal{M}$ to $\log\mathcal{M}+\mathrm{d}\log\mathcal{M}$,
where $\mathrm{d}\log\mathcal{M}$ is some log base 10 mass interval.
Adopting the GAMA stellar masses presented in \citet{Taylor2011},
we calculate our GSMF (and also our MSMFs below) via a direct summation
of stellar mass in bins of $0.1$ dex.

The GSMF may be described using a \citet{Schechter1976} function
whereby the number density, $\Phi\left(\log\mathcal{M}\right)\mathrm{d}\log\mathcal{M}$,
is given by
\begin{eqnarray}
\Phi\left(\log\mathcal{M}\right)\mathrm{d}\log\mathcal{M} & = & \ln(10)\cdot\phi^{*}10^{\log\left(\mathcal{M}/\mathcal{M}^{*}\right)\left(\alpha+1\right)}\nonumber \\
 &  & \times\exp\left(-10^{\log\left(\mathcal{M}/\mathcal{M}^{*}\right)}\right)\mathrm{d}\log\mathcal{M}\label{eq:schechter-log}
\end{eqnarray}
where $\mathcal{M}^{*}$ is the characteristic mass corresponding
to the position of the distinctive `knee' in the mass function. The
terms $\alpha$ and $\phi^{*}$ refer to the slope of the mass function
at the low mass end and the normalisation constant, respectively.
Several recent studies have previously measured the GSMF (e.g.; \citealp{Baldry2008a,Peng2010b,Baldry2012}),
and advocate the double Schechter form of the GSMF with a combined
knee ($\mathcal{M}^{*}$) for the global population. The double Schechter
function is simply given by $\Phi_{double}\left(\log\mathcal{M}\right)\mathrm{d}\log\mathcal{M}=\Phi_{1}+\Phi_{2}$,
where $\Phi_{1}$ and $\Phi_{2}$ refer to Equation \ref{eq:schechter-log}
above, albeit with separate slope parameters, $\alpha_{1}$ and $\alpha_{2}$,
and unique normalisation values, $\phi_{1}^{*}$ and $\phi_{2}^{*}$.
Both $\Phi_{1}$ and $\Phi_{2}$ share a common $\mathcal{M}^{*}$
parameter. The double Schechter function allows one to more accurately
model the distinctive bump observed in the GSMF about $\mathcal{M}^{*}$,
with one Schechter function dominant at stellar masses greater than
$\mathcal{M}^{*}$, and the second dominant otherwise. We adopt this
technique, opting to fit the GSMF with a double Schechter model%
\footnote{All Schechter functions are fit using the \noun{nlminb} routine in
R; a quasi-Newton algorithm based on the PORT routines that optimise
fitting in a similar sense to the Limited-memory Broyden-Fletcher-Goldfarb-Shanno
algorithm (LM-BFGS), with an extension to handle simple box constraints
on input variables (L-BFGS-B). The PORT documentation is available
at http://netlib.bell-labs.com/cm/cs/cstr/153.pdf%
}, however, we maintain a single Schechter model for the morphological-type
stellar mass functions (MSMFs hereafter) that constitute it.

\subsection{Morphological-Type Stellar Mass Functions}

\label{sub:msmf}
\begin{figure*}
\includegraphics[width=1\textwidth]{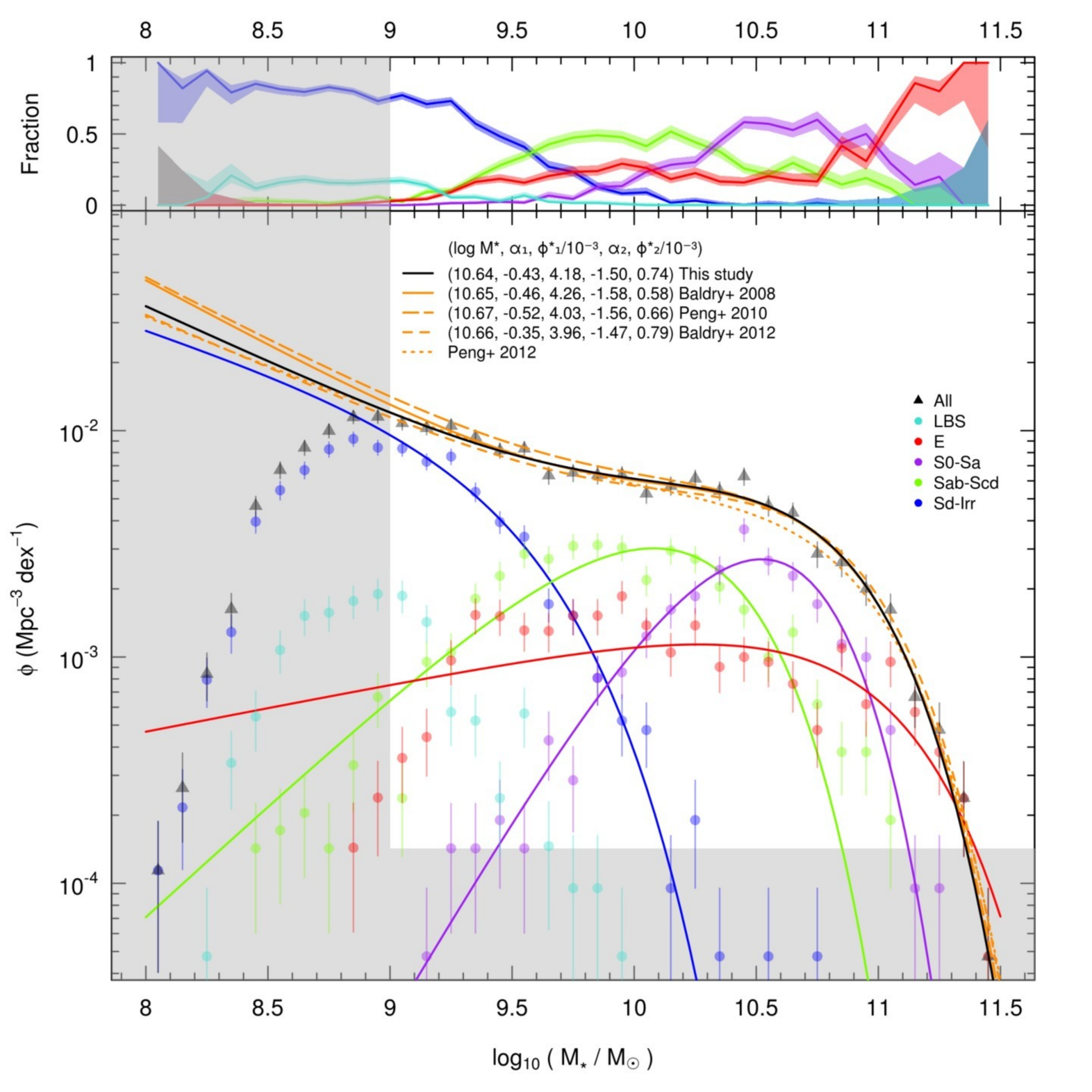}

\caption{\label{fig:massfunc}The Galaxy Stellar Mass Function and constituent
Morphological-Type Stellar Mass Functions as fit by double and single
Schechter functions respectively. Each galaxy population is labelled
and coloured according to the inset legend. The data is split into
mass bins of $0.1$ dex, with the error per bin assumed to be Poissonian
($\sqrt{n}$) in nature. Shaded grey areas ($\log\left(\mathcal{M}_{*}/\mathcal{M}_{\odot}\right)<9.0$
and number of galaxies $n\le3$) indicate those regions where data
has not been used in constraining the Schechter fits. Schechter fit
parameters for the GSMF in addition to fits from other studies are
also shown for reference. The upper panel shows the number fraction
of galaxies as a function of $V_{\mathrm{max}}$ weight corrected
stellar mass, calculated in mass bins identical to those in the lower
panel. Shaded coloured regions around each morphological-type fraction
line indicate the $\pm1\sigma$ confidence intervals, as calculated
using the \noun{qbeta} function \citep{Cameron2011}.}
\end{figure*}

Figure \ref{fig:massfunc} shows our GSMF and constituent
MSMFs for our volume and stellar mass limited GAMAnear$\mathcal{M}_{\mathrm{lim}}$
sample of $2,711$ systems. Stellar masses shown here have been $V_{\mathrm{max}}$
weight corrected where appropriate. The solid black line indicates
a double Schechter fit to the total GSMF, binned into mass bins of
$0.1$ dex, whilst the various orange lines show similar GSMF double
Schechter fits found in recent studies \citep{Baldry2008a,Peng2010b,Baldry2012,Peng2012}.
Note that we choose not to match to additional complementary studies,
such as that of Taylor et al. (2014, submitted) which divides their
sample into statistically defined `R' and `B' populations, or the
older yet still equally valid studies of \citet{Bell2003} and \citet{Baldry2004}.
This is for the sake of clarity alone, to avoid confusion within our
Figure \ref{fig:massfunc}. Solid coloured lines indicate single Schechter
fits to the constituent MSMFs, where colour relates to morphology
as indicated by the inset legend. Note that no Schechter fit to the
LBS population is shown, as there was not sufficient data to constrain
a Schechter function at this low mass end of the dataset. Shaded grey
areas ($\log\left(\mathcal{M}_{*}/\mathcal{M}_{\odot}\right)<9.0$
and number of galaxies $n\le3$) indicate those regions where data
has not been used in constraining the Schechter fits. Data points
below our lower mass limit are from the parent GAMAnear sample, and
are shown only for reference. Also consider that the GAMA dataset
exhibits a high level of spectroscopic completeness ($>98\%$) down
to its stated limiting apparent magnitude depth of $r=19.4$ mag \citep{Driver2011},
which precludes the possibility of severely impacting our measured
stellar mass functions. The upper panel of Figure \ref{fig:massfunc}
shows the number fraction of galaxies as a function of $V_{\mathrm{max}}$
weight corrected stellar mass, calculated in mass bins identical to
those in the lower panel. Shaded coloured regions around each morphological-type
fraction line within the upper panel indicate the $\pm1\sigma$ confidence
intervals, as calculated using the \noun{qbeta} function \citep{Cameron2011}.

We find our global GSMF in excellent agreement with the complementary
studies shown in Figure \ref{fig:massfunc}, exhibiting a comparable
$\mathcal{M}^{*}$ Schechter fit parameter at $\log\left(\mathcal{M}^{*}/\mathcal{M}_{\odot}\right)=10.64\pm0.07$,
and agreeing well within the errors. The high mass end of our sample
predominantly consists of spheroid dominated elliptical and S0-Sa
type galaxies. At intermediate masses below the global $\mathcal{M}^{*}$
value, the disk dominated Sab-Scd population dominates the stellar
mass budget, whilst at the low mass end of our dataset the Sd-Irr
and LBS populations are the most influential. It is apparent that
the latter LBS population is poorly sampled in this mass regime, with
the $\mathcal{M}^{*}$ parameter likely residing below our lower stellar
mass limit of $\log\left(\mathcal{M}_{*}/\mathcal{M}_{\odot}\right)=9.0$.
For this reason, we do not provide Schechter function fit parameters
to the LBS population in this study. We remind the reader that this
sample should be considered a field dominated sample, rather than
a cluster environment, as is evidenced by the dominance of Sd-Irr
and LBS type systems at the low mass end of our dataset. 

\begin{figure*}
\includegraphics[width=1\textwidth]{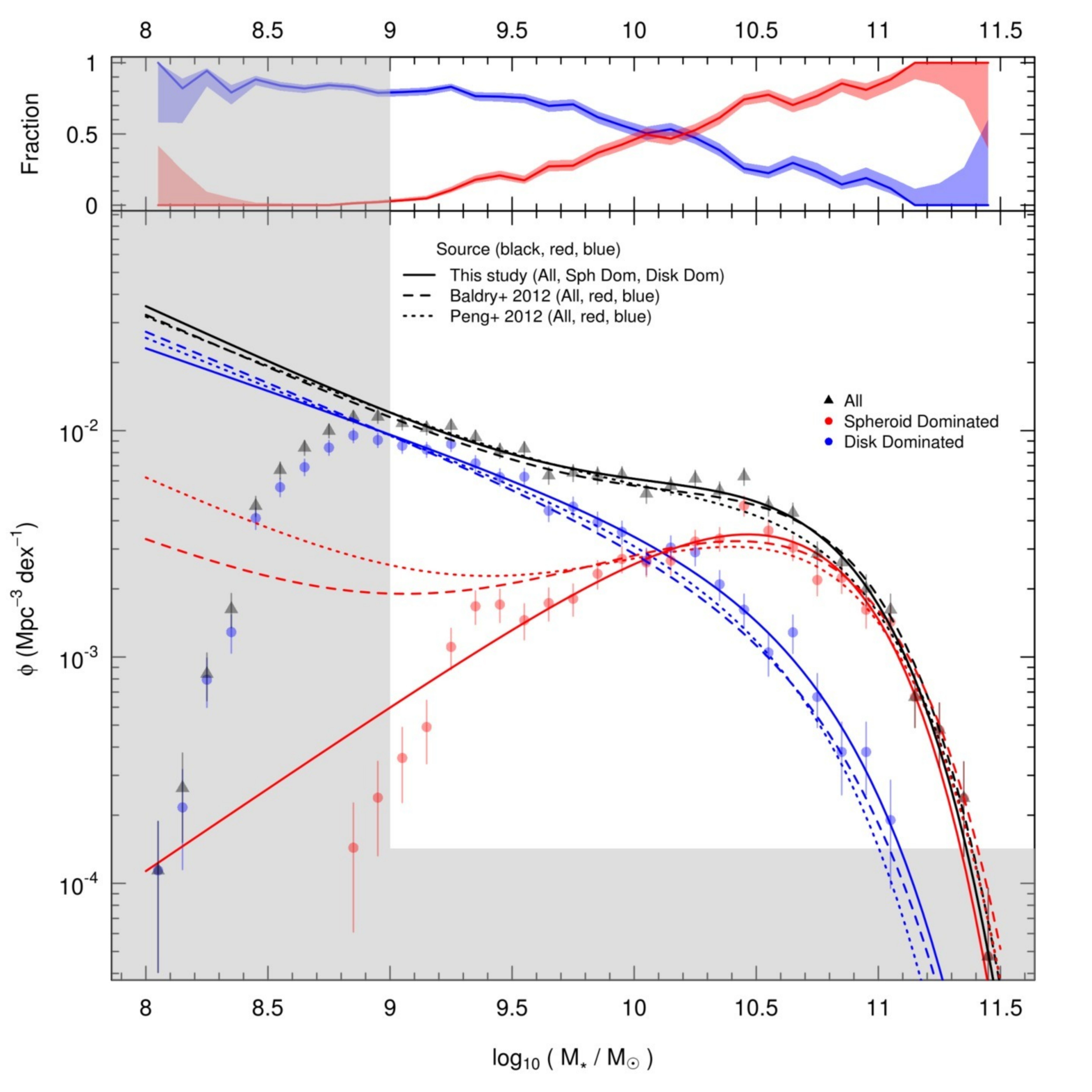}

\caption{\label{fig:massfunctype}As Figure \ref{fig:massfunc}, but for a
reduced grouping of morphological types, as indicated, which may broadly
be compared to early type and late type galaxies. Comparison Schechter
function fits for similar red and blue populations from \citet{Baldry2012}
and \citet{Peng2012} are also shown.}
\end{figure*}

Because of the uncertainty in our elliptical/S0-Sa division, and in
an attempt to group galaxies into structurally meaningful parent samples,
we now also combine our morphological types into two populations in
Figure \ref{fig:massfunctype} as indicated, namely: spheroid dominated
(E, S0-Sa) and disk dominated (Sab-Scd, Sd-Irr) galaxies, and similarly
fit these data with a single Schechter function. our recovered $\mathcal{M}^{*}$
Schechter fit parameters for our combined stellar mass functions are
remarkably similar to one another and to our total GSMF, with $\log\left(\mathcal{M}^{*}/\mathcal{M}_{\odot}\right)=10.60$
and $10.70$, respectively, supporting the notion that the combined
total galaxy stellar mass function is well described by a double Schechter
function comprised of two distinct components identified morphologically
here. Comparison Schechter function fits for a similar red and blue
population from \citet{Baldry2012} and \citet{Peng2012} are also
shown in Figure \ref{fig:massfunctype}. The \citeauthor{Peng2012}
GSMF is a summation of Schechter function fits to red/blue central
and red/blue satellite galaxies. We find our disk dominated population
in excellent agreement with the \citeauthor{Baldry2012} and \citeauthor{Peng2012}
blue populations, agreeing well at the lowest stellar masses, whilst
we find a slight surplus of stellar mass in disk dominated galaxies
at masses greater than $\mathcal{M}^{*}$. Our spheroid dominated
population similarly shows a good level of agreement at the most massive
end of our sample beyond $\mathcal{M}^{*}$, however; we do not find
the low mass turn up found in the comparison red populations. 

\begin{figure*}
\includegraphics[width=0.9\textwidth]{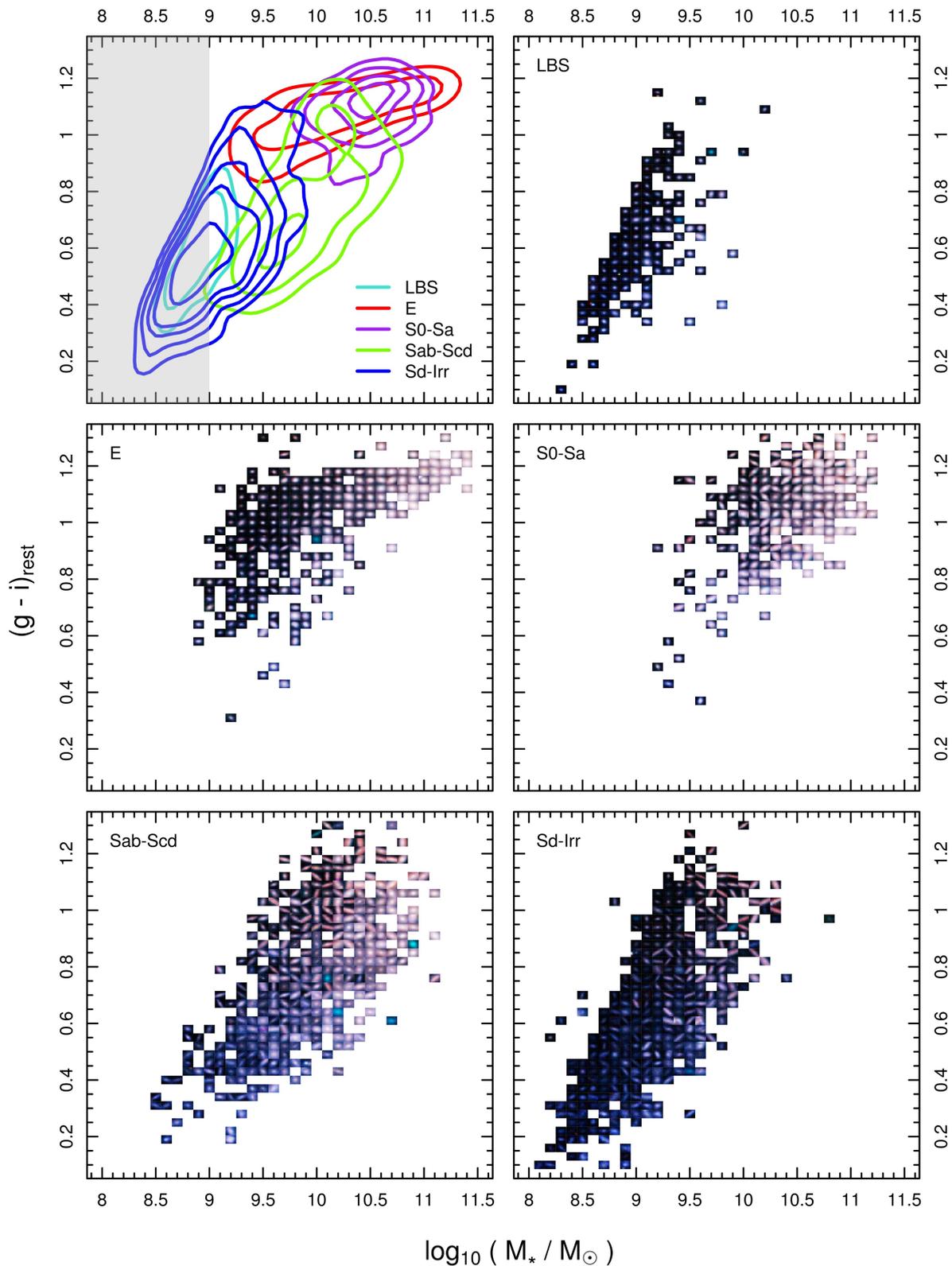}

\caption{\label{fig:masscolourgrid}Global rest frame colour, $(g-i){}_{\mathrm{rest}}$,
as a function of $V_{\mathrm{max}}$ weight corrected galaxy stellar
mass, $\log\left(\mathcal{M}_{*}/\mathcal{M}_{\odot}\right)$. The
top left panel displays several overlaid contour maps highlighting
the population density in colour-mass space for each of our morphological
types, as indicated by the inset legend. Contours represent $5$,
$10$, $20$, $40$ and $80$\% of the peak total population density.
The shaded grey area at $\log\left(\mathcal{M}_{*}/\mathcal{M}_{\odot}\right)<9.0$
represents the region in which our data becomes stellar mass incomplete.
The remaining panels represent a similar colour-mass space but for
each population in isolation, as labelled in the top left corner of
each panel. For each panel, we display a $3$-colour (RGB = $Hig$)
postage stamp image of a galaxy at each position within a grid of
bin sizes $0.1$ in stellar mass and $0.03$ in colour. Each postage
stamp is approximately $7''\times7''$ in size. Blank spaces indicate
a grid element where no galaxy of that morphological class exists.}

\end{figure*}

The cause of this discrepancy remains somewhat a mystery, and perhaps
rests with our choice of comparison samples. For example, \citet{Yang2009}
find no low-mass turn up for their red population across a stellar
mass regime comparable to that probed here, disagreeing with the studies
above, and highlighting apparent difficulties when dividing the galaxy
population by colour alone. Similarly, whilst \citet{Muzzin2013}
and \citet{Tomczak2014} do find a low-mass turn up in the stellar
mass function of quiescent galaxies when dividing the galaxy population
into quiescent/star-forming sub-populations, \citet{Omand2014} find
no noticeable low-mass turn up for their equivalent quiescent sample.
To expand on the relation between colour and stellar mass, Figure
\ref{fig:masscolourgrid} shows global rest frame colour, $(g-i)_{\mathrm{rest}}$,
as a function of galaxy stellar mass, $\log\left(\mathcal{M}_{*}/\mathcal{M}_{\odot}\right)$.
Our restframe colours are those derived concurrently with the stellar
masses from \citet{Taylor2011}, i.e., an SED fit to the GAMA galaxy
photometry. The top left panel displays several overlaid contour maps
highlighting the population density in colour-mass space for all of
our morphological types, as indicated by the inset legend, whereas
the remaining panels represent the same colour-mass space for each
population in isolation, as labelled in the top left corner of each
panel. For these latter panels, we display a $3$-colour (RGB = $Hig$)%
\footnote{Postage stamps of a peculiar turquoise colour indicate galaxies that
lie in a region where no near-infrared (UKIDSS-LAS) data was available
at the time of postage stamp creation, hence a missing red channel
in the creation of our $3$-colour images.%
} postage stamp image of a galaxy at each position within a grid of
bin size $0.1$ in stellar mass and $0.03$ in colour, where a blank
space indicates no galaxy of that morphological type exists. In total
in this figure, we display $1,245$ galaxies from our GAMAnear sample
($32.5\%$), with each postage stamp approximately $7''\times7''$
in size. We find the spheroid dominated elliptical and lenticular/early
type red sequence extending across a wide range of stellar masses,
$9.5<\log\left(\mathcal{M}_{*}/\mathcal{M}_{\odot}\right)<11.5$ with
a relatively small variation in global $(g-i)_{\mathrm{rest}}$ colour
across this range. As can also be seen in Figure \ref{fig:massfunc},
the elliptical population rarely dominates the stellar mass budget
in this range for any given stellar mass, except at the most massive
extreme of our sample ($\log\left(\mathcal{M}_{*}/\mathcal{M}_{\odot}\right)\sim11$).
At stellar masses below $\log\left(\mathcal{M}_{*}/\mathcal{M}_{\odot}\right)\approx10.2$,
first the disk dominated Sab-Scd population followed by the Sd-Irr
population provide a significant contamination fraction to the red
sequence. This contamination `break-point' is in good agreement with
that reported in Taylor et al. (2014, submitted), whereby the mean
colour of a statistically defined red population of galaxies jumps
by $\approx0.2$ mag, coupled with an increase in scatter, at $\log\left(\mathcal{M}_{*}/\mathcal{M}_{\odot}\right)\approx10.1$. 

One possible explanation behind this red sequence contamination becomes
apparent in the postage stamps for the Sab-Scd and Sd-Irr population
panels. In the colour regime $(g-i)_{\mathrm{rest}}>0.7$, a significant
fraction of disk dominated galaxies are observed highly inclined or
edge on. The reprocessing of galactic light as it travels through
a disk has the effect of reddening the resultant light due to the
effects of intrinsic dust attenuation, so therefore any photometric
estimate of the global colour will be biased redwards, in addition
to affecting other measured photometric properties (e.g.; \citealp{Pastrav2013a}).
This late type morphological contamination of the red sequence, effectively
a redistribution of stellar mass from the blue to the red population,
may perhaps be responsible for the observed turn up of the red population
stellar mass functions at the low mass end reported in, e.g., \citet{Baldry2012,Peng2012}.
Further, we posit that any such division of the local galaxy population
by (uncorrected) colour into a red sequence and blue cloud, such as
that adopted by, e.g., \citealp{Bell2003,Baldry2004} (a division
in colour-magnitude space) and \citealp{Peng2010b} (a division in
colour-stellar mass space), becomes increasingly meaningless at stellar
masses below $\log\left(\mathcal{M}_{*}/\mathcal{M}_{\odot}\right)\approx10.2$.

We stress however that colour is no more equivalent to spheroid/disk-dominated
than it is to quiescent/star-forming, slow/fast rotator, early-type/late-type
or metal rich/poor, to name but a few common bimodal galaxy identities.
Whilst significant overlaps may, and do, exist between these populations,
they do in fact measure distinct galaxy populations, and therefore
one may not always expect to recover a similar trend in, for example,
the observed stellar mass function. Also consider that our choice
to construct a spheroid-dominated sample from elliptical and S0-Sa
galaxies alone undoubtedly influences our recovered stellar mass functions.
We note that, should we choose to include the LBS population into
the spheroid-dominated population, we similarly recover a low-mass
turn up such as that observed in \citet{Baldry2012} and \citet{Peng2012}.
However, since LBS galaxies are notably blue, one might expect any
division by colour to bin LBS galaxies with our typically blue disk-dominated
systems, increasing the number density for the disk-dominated population
alone, and therefore not providing the required turn-up for spheroid-dominated
systems in the low mass regime. See Appendix \ref{app:lbs} for a
further discussion of the inclusion of the LBS population into our
spheroid dominated class.

Full Schechter fit parameters for both the double GSMF and constituent
MSMFs (both Hubble type morphologies and combined spheroid/disk dominated
populations) are shown in Tables \ref{tab:massfunc} and \ref{tab:massfuncmorph}
respectively. As previously noted, we do not provide Schechter fit
parameters for the LBS population. Errors on the $\rho_{\Sigma}$
parameter are propagated through from the stellar mass errors estimated
in \citet{Taylor2011}, typically of the order $\sim0.1$ dex. The
second set of errors for the $\log\mathcal{M}^{*}$, $\alpha$ and
$\phi^{*}$ parameters represent one standard deviation as derived
from comparable Schechter function fits to each individual observers
data set alone, giving an indication of classification agreement between
all three observers. All other errors provided in both tables are
estimated from jackknifed resampling using the relation $\sigma^{2}=\frac{N-1}{N}\sum_{i=1}^{N}\left(x_{j}-x\right)^{2}$,
where $x$ is the best fit parameter, $x_{j}$ is the best fit parameter
as given from a jackknife resampled variant of the data set and $N$
represents the number of jackknife volumes (we adopt $N=10$). 

The double Schechter GSMF provides a good fit to the bimodal form
of the total population, with a goodness of fit parameter of $\chi^{2}/\nu=1.12$
(a $\chi^{2}$ $p$-value of $p=0.33$, with $\chi^{2}=21.2$ and
$k=19$ degrees of freedom; i.e., we have insufficient evidence to
reject our fitted model). As can be inferred from Figure \ref{fig:massfunc}
and the gradient of the elliptical population in Table \ref{tab:massfuncmorph},
the initial high-mass peak primarily consists of S0-Sa galaxies, with
some small contribution from elliptical galaxies. Ellipticals appear
to exist uniformly across a wide range of masses. Our fitted Schechter
function to the elliptical population appears to be a relatively poor
fit to the data, as evidenced by the goodness of fit parameter and
confidence intervals quoted in Table \ref{tab:massfuncmorph}, able
to capture the high mass turnover about $\mathcal{M}^{*}$ but partially
underestimating the number counts at lower stellar masses. No doubt
this discrepancy is caused by the inflexibility of the Schechter function
in fitting a population that is uniformly distributed in number density
such as this. Similarly, the goodness of fit parameter for the Sab-Scd
population is quite poor. From Figure \ref{fig:massfunc} we see that
this discrepancy occurs at the high mass end, above $\mathcal{M}^{*}$,
with an unexpected surplus of galaxies and a departure from the Schechter
fit at $\log\left(\mathcal{M}^{*}/\mathcal{M}_{\odot}\right)\sim11$.
This could be evidence of perhaps spheroid dominated (elliptical,
lenticular or early-type spiral) contamination of the Sab-Scd population
in this regime. In addition, errors arising from observer disagreement
place a significant level of uncertainty on Schechter fit parameters
to our Sd-Irr population. This implies that perhaps these data are
not of a sufficient depth to fully measure the characteristic turn-over
in the Sd-Irr stellar mass function. Visual classification error for
the remaining morphological types remains minimal however, typically
of the order of or less than the quoted standard errors. We find that
our recovered $\mathcal{M}^{*}$ parameter for our constituent MSMFs
decrease systematically from spheroid dominated to disk dominated
galaxies; for E, S0-Sa, Sab-Scd and Sd-Irr type galaxies we find $\log\left(\mathcal{M}^{*}/\mathcal{M}_{\odot}\right)=10.94$,
$10.25$, $10.09$, $9.57$. 

Our combined spheroid dominated and disk dominated single Schechter
fits provide an excellent description of the spheroid and disk dominated
galaxy populations. The goodness of fit estimators both indicate the
Schechter model is able to adequately and accurately reproduce the
distribution observed in the data, whilst the quoted errors, both
standard and visual, remain low. Further, we note that the recovered
Schechter fit parameters to our spheroid-dominated and disk-dominated
populations: $\log\mathcal{M}^{*}=10.60$, $10.70$; $\alpha=-0.27$,
$-1.37$, and; $\phi^{*}=3.96$, $0.98$, respectively, are in good
agreement with those found for our double Schechter fit to the total
population: $\log\mathcal{M}^{*}=10.64$, $10.64$; $\alpha=-0.43$,
$-1.50$; $\phi^{*}=4.18$, $0.74$. The apparent self-similarity
between these two sets of recovered parameters supports the notion
that our division of the GSMF into spheroid-dominated and disk-dominated
sub-populations is indeed physically meaningful. By dividing galaxies
according to their dominant structural component, we have been able
to naturally recover the fundamental parameters which best describe
the full stellar mass distribution of galaxies in the local Universe.

\renewcommand{\arraystretch}{1.5}
\setlength{\tabcolsep}{6pt}

\begin{table*}
\begin{centering}
\begin{tabular}{cccccccc}
\hline 
$\log\left(\mathcal{M}^{*}/\mathcal{M}_{\odot}\right)$ & $\alpha_{1}$ & $\phi_{1}^{*}/10^{-3}$ & $\alpha_{2}$ & $\phi_{2}^{*}/10^{-3}$ & $\chi^{2}/\nu$ & $\delta_{\phi}/10^{7}$ & $\delta_{\Sigma}/10^{7}$\tabularnewline
 &  & ($\mathrm{dex}^{-1}\mathrm{Mpc}^{-3}$) &  & ($\mathrm{dex}^{-1}\mathrm{Mpc}^{-3}$) &  & ($\mathcal{M}_{\odot}\mathrm{Mpc^{-3}}$) & ($\mathcal{M}_{\odot}\mathrm{Mpc^{-3}}$) \tabularnewline
\hline 
$10.64\pm0.07$ & $-0.43\pm0.35$ & $4.18\pm1.52$ & $-1.50\pm0.22$ & $0.74\pm1.13$ & $1.12$ & $22.07\pm7.91$ & $21.88_{-5.22}^{+6.90}$ \tabularnewline
\hline 
\end{tabular}
\par\end{centering}

\caption{\label{tab:massfunc}Double Schechter stellar mass function fit parameters
for the total GSMF as shown in Figure \ref{fig:massfunc}. From left
to right, columns are: the shared knee in the Schechter function ($\mathcal{M}^{*}$);
the primary slope of the faint end of the Schechter function ($\alpha_{1}$);
the primary normalisation constant for the Schechter function ($\phi_{1}^{*}$);
the secondary slope of the faint end of the Schechter function ($\alpha_{2}$);
the secondary normalisation constant for the Schechter function ($\phi_{2}^{*}$);
the $\chi^{2}$ goodness of fit parameter ($\chi^{2}/\nu$); the stellar
mass density implied in the usual way via the fitted Schechter function
{[}$\rho_{\phi}=\sum_{i=1}^{N}\phi_{i}^{*}\mathcal{M}^{*}\Gamma\left(\alpha_{i}+2\right)${]};
the stellar mass density calculated via the direct summation of the
stellar masses from the individual galaxies {[}$\rho_{\Sigma}=\frac{1}{V}\sum_{i=1}^{N}\mathcal{M}_{*}${]}.
The double Schechter function is fit to $N=24$ data bins with $n=5$
fitted parameters, therefore, the number of degrees of freedom for
this fit is given by $\nu=N-n=19$.}
\end{table*}

\renewcommand{\arraystretch}{1}

\renewcommand{\arraystretch}{1.5}
\setlength{\tabcolsep}{6pt}

\begin{table*}
\begin{centering}
\begin{tabular}{ccccccc}
\hline 
Population & $\log\left(\mathcal{M}^{*}/\mathcal{M}_{\odot}\right)$ & $\alpha$ & $\phi^{*}/10^{-3}$ & $\chi^{2}/\nu$ & $\rho_{\phi}/10^{7}$ & $\rho_{\Sigma}/10^{7}$\tabularnewline
 &  &  & ($\mathrm{dex}^{-1}\mathrm{Mpc}^{-3}$) &  & ($\mathcal{M}_{\odot}\mathrm{Mpc^{-3}}$) & ($\mathcal{M}_{\odot}\mathrm{Mpc^{-3}}$) \tabularnewline
\hline 
E 		  & $10.94\pm0.10\pm0.18$ & $-0.79\pm0.13\pm0.23$ & $0.85\pm0.27\pm0.49$ & $3.10$ & $6.81\pm1.47$ & $7.46_{-1.81}^{+2.40}$ \tabularnewline
S0-Sa 	  & $10.25\pm0.07\pm0.03$ & $0.87\pm0.23\pm0.15$ & $2.38\pm0.83\pm0.27$ & $1.11$ & $7.53\pm2.08$ & $7.98_{-1.90}^{+2.51}$ \tabularnewline
Sab-Scd    & $10.09\pm0.15\pm0.09$ & $-0.01\pm0.31\pm0.26$ & $3.57\pm0.81\pm0.63$ & $4.07$ & $4.34\pm1.62$ & $5.18_{-1.20}^{+1.57}$ \tabularnewline
Sd-Irr 	  & $9.57\pm0.17\pm1.30$ & $-1.36\pm0.29\pm0.80$ & $3.40\pm2.07\pm4.29$ & $1.66$ & $1.77\pm1.10$ & $1.13_{-0.28}^{+0.38}$ \tabularnewline
\hline
Spheroid Dominated     & $10.60\pm0.05\pm0.08$ & $-0.27\pm0.16\pm0.20$ & $3.96\pm1.05\pm0.37$ & $1.59$ & $14.49\pm3.92$ & $15.44_{-3.71}^{+4.91}$ \tabularnewline
Disk Dominated & $10.70\pm0.23\pm0.07$ & $-1.37\pm0.11\pm0.04$ & $0.98\pm0.42\pm0.14$ & $0.90$ & $7.08\pm3.91$ & $6.31_{-1.48}^{+1.95}$ \tabularnewline
\hline 
\end{tabular}
\par\end{centering}

\caption{\label{tab:massfuncmorph}As Table \ref{tab:massfunc}, but single
Schechter stellar mass function fit parameters for the Morphological-Type
Stellar Mass Functions (MSMFs) in Figures \ref{fig:massfunc} and
\ref{fig:massfunctype}. The second set of confidence intervals for
the $\log\mathcal{M}^{*}$, $\alpha$ and $\phi^{*}$ parameters indicate
one standard deviation as determined from single-Schechter fits to
each individual observers data sets. These single Schechter functions
are fit to $N=24$ data bins with $n=3$ fitted parameters, therefore,
the number of degrees of freedom for this fit is given by $\nu=N-n=21$.}
\end{table*}

\renewcommand{\arraystretch}{1}

\section{Conclusions}

We have analysed a morphologically classified sample of $2,711$ galaxies
selected from the GAMA survey by virtue of their redshift range ($0.025<z<0.06$)
and global stellar mass ($\log\left(\mathcal{M}_{*}/\mathcal{M}_{\odot}\right)>9.0$).
Each galaxy is classified into either elliptical (E), spheroid-dominated
lenticular and early-type spiral (S0-Sa), intermediate/late-type spiral
(Sab-Scd) and a disk-dominated or irregular (Sd-Irr) class. Within
this local sample, we find approximate stellar mass proportions for
E : S0-Sa : Sab-Scd : Sd-Irr of $34$ : $37$ : $24$ : $5$, acknowledging
a potential cross-contamination between our elliptical and S0-Sa classes.
We find that colour and mass cuts do not trivially recover Hubble
type classifications and advocate against using `red' and `blue' terminology
interchangeably with `early' and `late', or `spheroid dominated' and
`disk dominated' as these are clearly very different distinctions.
Grouping by the dominant structural component, spheroid or disk, we
further find that approximately $71{}_{-4}^{+3}$\% of the stellar
mass is currently found within spheroid dominated elliptical and S0-Sa
type galaxies, with $29{}_{-3}^{+4}$\% residing in disk dominated
Sab-Scd and Sd-Irr systems. Adopting reasonable bulge-to-total values
(e.g., \citealp{Graham2008b}) implies that approximately half the
stellar mass today resides in spheroidal structures, with the remaining
half within disk-like structures, in-line with previous studies (see
\citealp{Driver2007a,Driver2007b,Gadotti2009,Tasca2011}). 

The total galaxy stellar mass function for our sample is well described
by a double Schechter function with parameters $\mathcal{M}^{*}=10^{10.64}\mathcal{M}_{\odot}$,
$\alpha_{1}=-0.43$, $\phi_{1}^{*}=4.18\;\mathrm{dex}^{-1}\mathrm{Mpc}^{-3}$,
$\alpha_{2}=-1.50$ and $\phi_{2}^{*}=0.74\;\mathrm{dex}^{-1}\mathrm{Mpc}^{-3}$.
The constituent morphological-type stellar mass functions are well
sampled above our lower stellar mass limit, with the exception of
the little blue spheroid population, which remains incomplete down
to $\log\left(\mathcal{M}_{*}/\mathcal{M}_{\odot}\right)\sim9.0$.
Each morphological-type stellar mass function is adequately described
by a single Schechter function (Figure \ref{fig:massfunc}), with
a notable underestimation of the number density of elliptical galaxies
at low stellar masses ($\log\left(\mathcal{M}_{*}/\mathcal{M}_{\odot}\right)<10$),
and an underestimation of our Sab-Scd population number density at
high stellar masses ($\log\left(\mathcal{M}_{*}/\mathcal{M}_{\odot}\right)\sim11$).
We find our recovered $\mathcal{M}^{*}$ for these morphological-type
stellar mass functions decreases systematically from spheroid dominated
to disk dominated galaxies, i.e.; for E, S0-Sa, Sab-Scd and Sd-Irr
type galaxies we find $\log\left(\mathcal{M}^{*}/\mathcal{M}_{\odot}\right)=10.94$,
$10.25$, $10.09$, $9.57$, respectively. 

Our combined spheroid dominated and disk dominated stellar mass functions
are each well described by a single Schechter function (Figure \ref{fig:massfunctype}).
Interestingly, our recovered $\mathcal{M}^{*}$ parameters for our
combined spheroid dominated and disk dominated stellar mass functions
are remarkably similar to one another, in addition to our total galaxy
stellar mass function, with $\log\left(\mathcal{M}^{*}/\mathcal{M}_{\odot}\right)=10.60$
and $10.70$ respectively, as compared with $\log\left(\mathcal{M}^{*}/\mathcal{M}_{\odot}\right)=10.64$.
We also find a good level of agreement between our spheroid and disk-dominated
populations and the total galaxy stellar mass function for the additional
Schechter fit parameters, $\alpha$ and $\phi^{*}$. That these two
sets of values should arise naturally from the data supports the notion
that the combined total galaxy stellar mass function is indeed comprised
of two complementary, yet distinct, sub-populations, each best described
according to their dominant structural component. We find that the
discrepancy between our spheroid dominated stellar mass function and
the comparison red sequence stellar mass functions of \citet{Baldry2012}
and \citet{Peng2012} at the low mass end of our sample can potentially
be attributed to late type contamination of the red sequence (Figure
\ref{fig:masscolourgrid}), although we note that a division of the
local galaxy population by colour may not easily be comparable to
a division by dominant structural component; nor should it. In addition,
the inclusion of the LBS population into the spheroid dominated class
acts to remove the observed low-mass discrepancy, however; it is not
clear that this inclusion is desired. Therefore, in conclusion, our
campaign of robust morphological classification shows that the local
galaxy stellar mass function is adequately described by a double Schechter
function comprised of two distinct populations: spheroid dominated
and disk dominated galaxies.

\section*{Acknowledgements}

This work was supported by the Austrian Science Foundation FWF under
grant P23946. AWG was supported under the Australian Research Council\textquoteright{}s
funding scheme FT110100263. GAMA is a joint European-Australasian
project based around a spectroscopic campaign using the Anglo-Australian
Telescope. The GAMA input catalogue is based on data taken from the
Sloan Digital Sky Survey and the UKIRT Infrared Deep Sky Survey. Complementary
imaging of the GAMA regions is being obtained by a number of independent
survey programs including GALEX MIS, VST KiDS, VISTA VIKING, WISE,
Herschel-ATLAS, GMRT and ASKAP providing UV to radio coverage. GAMA
is funded by the STFC (UK), the ARC (Australia), the AAO, and the
participating institutions. The GAMA website is http://www.gama-survey.org/.

\bibliographystyle{mn2e}
\bibliography{biblib}

\begin{thebibliography}{96}
\expandafter\ifx\csname natexlab\endcsname\relax\def\natexlab#1{#1}\fi

\bibitem[{{Abazajian} {et~al}\mbox{.}(2009){Abazajian}, {Adelman-McCarthy},
  {Ag{\"u}eros}, {Allam}, {Allende Prieto}, {An}, {Anderson}, {Anderson},
  {Annis}, {Bahcall}, \& et~al.}]{Abazajian2009}
{Abazajian} K.~N. {et~al.}, 2009, \apjs, 182, 543

\bibitem[{{Allen} {et~al}\mbox{.}(2006){Allen}, {Driver}, {Graham}, {Cameron},
  {Liske}, \& {de Propris}}]{Allen2006}
{Allen} P.~D., {Driver} S.~P., {Graham} A.~W., {Cameron} E., {Liske} J., {de
  Propris} R., 2006, \mnras, 371, 2

\bibitem[{{Baldry} {et~al}\mbox{.}(2006){Baldry}, {Balogh}, {Bower},
  {Glazebrook}, {Nichol}, {Bamford}, \& {Budavari}}]{Baldry2006}
{Baldry} I.~K., {Balogh} M.~L., {Bower} R.~G., {Glazebrook} K., {Nichol} R.~C.,
  {Bamford} S.~P., {Budavari} T., 2006, \mnras, 373, 469

\bibitem[{{Baldry} {et~al}\mbox{.}(2012){Baldry}, {Driver}, {Loveday},
  {Taylor}, {Kelvin}, {Liske}, {Norberg}, {Robotham}, {Brough}, {Hopkins},
  {Bamford}, {Peacock}, {Bland-Hawthorn}, {Conselice}, {Croom}, {Jones},
  {Parkinson}, {Popescu}, {Prescott}, {Sharp}, \& {Tuffs}}]{Baldry2012}
{Baldry} I.~K. {et~al.}, 2012, \mnras, 421, 621

\bibitem[{{Baldry} {et~al}\mbox{.}(2004){Baldry}, {Glazebrook}, {Brinkmann},
  {Ivezi{\'c}}, {Lupton}, {Nichol}, \& {Szalay}}]{Baldry2004}
{Baldry} I.~K., {Glazebrook} K., {Brinkmann} J., {Ivezi{\'c}} {\v Z}., {Lupton}
  R.~H., {Nichol} R.~C., {Szalay} A.~S., 2004, \apj, 600, 681

\bibitem[{{Baldry} {et~al}\mbox{.}(2008){Baldry}, {Glazebrook}, \&
  {Driver}}]{Baldry2008a}
{Baldry} I.~K., {Glazebrook} K., {Driver} S.~P., 2008, \mnras, 388, 945

\bibitem[{{Baldry} {et~al}\mbox{.}(2010){Baldry}, {Robotham}, {Hill}, {Driver},
  {Liske}, {Norberg}, {Bamford}, {Hopkins}, {Loveday}, {Peacock}, {Cameron},
  {Croom}, {Cross}, {Doyle}, {Dye}, {Frenk}, {Jones}, {van Kampen}, {Kelvin},
  {Nichol}, {Parkinson}, {Popescu}, {Prescott}, {Sharp}, {Sutherland},
  {Thomas}, \& {Tuffs}}]{Baldry2010}
{Baldry} I.~K. {et~al.}, 2010, \mnras, 404, 86

\bibitem[{{Bamford} {et~al}\mbox{.}(2009){Bamford}, {Nichol}, {Baldry}, {Land},
  {Lintott}, {Schawinski}, {Slosar}, {Szalay}, {Thomas}, {Torki}, {Andreescu},
  {Edmondson}, {Miller}, {Murray}, {Raddick}, \& {Vandenberg}}]{Bamford2009}
{Bamford} S.~P. {et~al.}, 2009, \mnras, 393, 1324

\bibitem[{{Behroozi} {et~al}\mbox{.}(2013){Behroozi}, {Wechsler}, \&
  {Conroy}}]{Behroozi2013}
{Behroozi} P.~S., {Wechsler} R.~H., {Conroy} C., 2013, \apj, 770, 57

\bibitem[{{Bell} {et~al}\mbox{.}(2003){Bell}, {McIntosh}, {Katz}, \&
  {Weinberg}}]{Bell2003}
{Bell} E.~F., {McIntosh} D.~H., {Katz} N., {Weinberg} M.~D., 2003, \apjs, 149,
  289

\bibitem[{{Blanton} {et~al}\mbox{.}(2005){Blanton}, {Eisenstein}, {Hogg},
  {Schlegel}, \& {Brinkmann}}]{Blanton2005a}
{Blanton} M.~R., {Eisenstein} D., {Hogg} D.~W., {Schlegel} D.~J., {Brinkmann}
  J., 2005, \apj, 629, 143

\bibitem[{{Bolzonella} {et~al}\mbox{.}(2010){Bolzonella}, {Kova{\v c}},
  {Pozzetti}, {Zucca}, {Cucciati}, {Lilly}, {Peng}, {Iovino}, {Zamorani},
  {Vergani}, {Tasca}, {Lamareille}, {Oesch}, {Caputi}, {Kampczyk}, {Bardelli},
  {Maier}, {Abbas}, {Knobel}, {Scodeggio}, {Carollo}, {Contini}, {Kneib}, {Le
  F{\`e}vre}, {Mainieri}, {Renzini}, {Bongiorno}, {Coppa}, {de la Torre}, {de
  Ravel}, {Franzetti}, {Garilli}, {Le Borgne}, {Le Brun}, {Mignoli},
  {Pell{\'o}}, {Perez-Montero}, {Ricciardelli}, {Silverman}, {Tanaka},
  {Tresse}, {Bottini}, {Cappi}, {Cassata}, {Cimatti}, {Guzzo}, {Koekemoer},
  {Leauthaud}, {Maccagni}, {Marinoni}, {McCracken}, {Memeo}, {Meneux},
  {Porciani}, {Scaramella}, {Aussel}, {Capak}, {Halliday}, {Ilbert},
  {Kartaltepe}, {Salvato}, {Sanders}, {Scarlata}, {Scoville}, {Taniguchi}, \&
  {Thompson}}]{Bolzonella2010}
{Bolzonella} M. {et~al.}, 2010, \aap, 524, A76

\bibitem[{{Bruzual} \& {Charlot}(2003)}]{Bruzual2003}
{Bruzual} G., {Charlot} S., 2003, \mnras, 344, 1000

\bibitem[{{Bundy} {et~al}\mbox{.}(2010){Bundy}, {Scarlata}, {Carollo}, {Ellis},
  {Drory}, {Hopkins}, {Salvato}, {Leauthaud}, {Koekemoer}, {Murray}, {Ilbert},
  {Oesch}, {Ma}, {Capak}, {Pozzetti}, \& {Scoville}}]{Bundy2010}
{Bundy} K. {et~al.}, 2010, \apj, 719, 1969

\bibitem[{{Calzetti} {et~al}\mbox{.}(2000){Calzetti}, {Armus}, {Bohlin},
  {Kinney}, {Koornneef}, \& {Storchi-Bergmann}}]{Calzetti2000}
{Calzetti} D., {Armus} L., {Bohlin} R.~C., {Kinney} A.~L., {Koornneef} J.,
  {Storchi-Bergmann} T., 2000, \apj, 533, 682

\bibitem[{{Cameron}(2011)}]{Cameron2011}
{Cameron} E., 2011, \pasa, 28, 128

\bibitem[{{Caon} {et~al}\mbox{.}(1993){Caon}, {Capaccioli}, \&
  {D'Onofrio}}]{Caon1993}
{Caon} N., {Capaccioli} M., {D'Onofrio} M., 1993, \mnras, 265, 1013

\bibitem[{{Cappellari} {et~al}\mbox{.}(2011{\natexlab{a}}){Cappellari},
  {Emsellem}, {Krajnovi{\'c}}, {McDermid}, {Scott}, {Verdoes Kleijn}, {Young},
  {Alatalo}, {Bacon}, {Blitz}, {Bois}, {Bournaud}, {Bureau}, {Davies}, {Davis},
  {de Zeeuw}, {Duc}, {Khochfar}, {Kuntschner}, {Lablanche}, {Morganti}, {Naab},
  {Oosterloo}, {Sarzi}, {Serra}, \& {Weijmans}}]{Cappellari2011a}
{Cappellari} M. {et~al.}, 2011{\natexlab{a}}, \mnras, 413, 813

\bibitem[{{Cappellari} {et~al}\mbox{.}(2011{\natexlab{b}}){Cappellari},
  {Emsellem}, {Krajnovi{\'c}}, {McDermid}, {Serra}, {Alatalo}, {Blitz}, {Bois},
  {Bournaud}, {Bureau}, {Davies}, {Davis}, {de Zeeuw}, {Khochfar},
  {Kuntschner}, {Lablanche}, {Morganti}, {Naab}, {Oosterloo}, {Sarzi}, {Scott},
  {Weijmans}, \& {Young}}]{Cappellari2011b}
{Cappellari} M. {et~al.}, 2011{\natexlab{b}}, \mnras, 416, 1680

\bibitem[{{Cappellari} {et~al}\mbox{.}(2013){Cappellari}, {McDermid},
  {Alatalo}, {Blitz}, {Bois}, {Bournaud}, {Bureau}, {Crocker}, {Davies},
  {Davis}, {de Zeeuw}, {Duc}, {Emsellem}, {Khochfar}, {Krajnovi{\'c}},
  {Kuntschner}, {Morganti}, {Naab}, {Oosterloo}, {Sarzi}, {Scott}, {Serra},
  {Weijmans}, \& {Young}}]{Cappellari2013}
{Cappellari} M. {et~al.}, 2013, \mnras, 432, 1862

\bibitem[{{Chabrier}(2003)}]{Chabrier2003}
{Chabrier} G., 2003, \pasp, 115, 763

\bibitem[{{Conselice}(2006)}]{Conselice2006b}
{Conselice} C.~J., 2006, \mnras, 373, 1389

\bibitem[{{Cook} {et~al}\mbox{.}(2010){Cook}, {Evoli}, {Barausse}, {Granato},
  \& {Lapi}}]{Cook2010a}
{Cook} M., {Evoli} C., {Barausse} E., {Granato} G.~L., {Lapi} A., 2010, \mnras,
  402, 941

\bibitem[{{Cook} {et~al}\mbox{.}(2009){Cook}, {Lapi}, \& {Granato}}]{Cook2009}
{Cook} M., {Lapi} A., {Granato} G.~L., 2009, \mnras, 397, 534

\bibitem[{{Cooper} {et~al}\mbox{.}(2012){Cooper}, {Griffith}, {Newman}, {Coil},
  {Davis}, {Dutton}, {Faber}, {Guhathakurta}, {Koo}, {Lotz}, {Weiner},
  {Willmer}, \& {Yan}}]{Cooper2012}
{Cooper} M.~C. {et~al.}, 2012, \mnras, 419, 3018

\bibitem[{{Davies} {et~al}\mbox{.}(1983){Davies}, {Efstathiou}, {Fall},
  {Illingworth}, \& {Schechter}}]{Davies1983a}
{Davies} R.~L., {Efstathiou} G., {Fall} S.~M., {Illingworth} G., {Schechter}
  P.~L., 1983, \apj, 266, 41

\bibitem[{{Davies} \& {Illingworth}(1983)}]{Davies1983b}
{Davies} R.~L., {Illingworth} G., 1983, \apj, 266, 516

\bibitem[{{De Lucia} \& {Blaizot}(2007)}]{DeLucia2007}
{De Lucia} G., {Blaizot} J., 2007, \mnras, 375, 2

\bibitem[{{De Lucia} {et~al}\mbox{.}(2004){De Lucia}, {Kauffmann}, \&
  {White}}]{DeLucia2004}
{De Lucia} G., {Kauffmann} G., {White} S.~D.~M., 2004, \mnras, 349, 1101

\bibitem[{{Debattista} {et~al}\mbox{.}(2006){Debattista}, {Mayer}, {Carollo},
  {Moore}, {Wadsley}, \& {Quinn}}]{Debattista2006}
{Debattista} V.~P., {Mayer} L., {Carollo} C.~M., {Moore} B., {Wadsley} J.,
  {Quinn} T., 2006, \apj, 645, 209

\bibitem[{{D'Onofrio} {et~al}\mbox{.}(1995){D'Onofrio}, {Zaggia}, {Longo},
  {Caon}, \& {Capaccioli}}]{D'Onofrio1995}
{D'Onofrio} M., {Zaggia} S.~R., {Longo} G., {Caon} N., {Capaccioli} M., 1995,
  \aap, 296, 319

\bibitem[{{Driver} {et~al}\mbox{.}(2006){Driver}, {Allen}, {Graham}, {Cameron},
  {Liske}, {Ellis}, {Cross}, {De Propris}, {Phillipps}, \&
  {Couch}}]{Driver2006}
{Driver} S.~P. {et~al.}, 2006, \mnras, 368, 414

\bibitem[{{Driver} {et~al}\mbox{.}(2007{\natexlab{a}}){Driver}, {Allen},
  {Liske}, \& {Graham}}]{Driver2007a}
{Driver} S.~P., {Allen} P.~D., {Liske} J., {Graham} A.~W., 2007{\natexlab{a}},
  \apjl, 657, L85

\bibitem[{{Driver} {et~al}\mbox{.}(2011){Driver}, {Hill}, {Kelvin}, {Robotham},
  {Liske}, {Norberg}, {Baldry}, {Bamford}, {Hopkins}, {Loveday}, {Peacock},
  {Andrae}, {Bland-Hawthorn}, {Brough}, {Brown}, {Cameron}, {Ching}, {Colless},
  {Conselice}, {Croom}, {Cross}, {de Propris}, {Dye}, {Drinkwater}, {Ellis},
  {Graham}, {Grootes}, {Gunawardhana}, {Jones}, {van Kampen}, {Maraston},
  {Nichol}, {Parkinson}, {Phillipps}, {Pimbblet}, {Popescu}, {Prescott},
  {Roseboom}, {Sadler}, {Sansom}, {Sharp}, {Smith}, {Taylor}, {Thomas},
  {Tuffs}, {Wijesinghe}, {Dunne}, {Frenk}, {Jarvis}, {Madore}, {Meyer},
  {Seibert}, {Staveley-Smith}, {Sutherland}, \& {Warren}}]{Driver2011}
{Driver} S.~P. {et~al.}, 2011, \mnras, 413, 971

\bibitem[{{Driver} {et~al}\mbox{.}(2005){Driver}, {Liske}, {Cross}, {De
  Propris}, \& {Allen}}]{Driver2005}
{Driver} S.~P., {Liske} J., {Cross} N.~J.~G., {De Propris} R., {Allen} P.~D.,
  2005, \mnras, 360, 81

\bibitem[{{Driver} {et~al}\mbox{.}(2009){Driver}, {Norberg}, {Baldry},
  {Bamford}, {Hopkins}, {Liske}, {Loveday}, {Peacock}, {Hill}, {Kelvin},
  {Robotham}, {Cross}, {Parkinson}, {Prescott}, {Conselice}, {Dunne}, {Brough},
  {Jones}, {Sharp}, {van Kampen}, {Oliver}, {Roseboom}, {Bland-Hawthorn},
  {Croom}, {Ellis}, {Cameron}, {Cole}, {Frenk}, {Couch}, {Graham}, {Proctor},
  {De Propris}, {Doyle}, {Edmondson}, {Nichol}, {Thomas}, {Eales}, {Jarvis},
  {Kuijken}, {Lahav}, {Madore}, {Seibert}, {Meyer}, {Staveley-Smith},
  {Phillipps}, {Popescu}, {Sansom}, {Sutherland}, {Tuffs}, \&
  {Warren}}]{Driver2009}
{Driver} S.~P. {et~al.}, 2009, Astronomy and Geophysics, 50, 050000

\bibitem[{{Driver} {et~al}\mbox{.}(2007{\natexlab{b}}){Driver}, {Popescu},
  {Tuffs}, {Liske}, {Graham}, {Allen}, \& {de Propris}}]{Driver2007b}
{Driver} S.~P., {Popescu} C.~C., {Tuffs} R.~J., {Liske} J., {Graham} A.~W.,
  {Allen} P.~D., {de Propris} R., 2007{\natexlab{b}}, \mnras, 379, 1022

\bibitem[{{Driver} {et~al}\mbox{.}(2013){Driver}, {Robotham}, {Bland-Hawthorn},
  {Brown}, {Hopkins}, {Liske}, {Phillipps}, \& {Wilkins}}]{Driver2013}
{Driver} S.~P., {Robotham} A.~S.~G., {Bland-Hawthorn} J., {Brown} M., {Hopkins}
  A., {Liske} J., {Phillipps} S., {Wilkins} S., 2013, \mnras, 430, 2622

\bibitem[{{Duc} {et~al}\mbox{.}(2011){Duc}, {Cuillandre}, {Serra},
  {Michel-Dansac}, {Ferriere}, {Alatalo}, {Blitz}, {Bois}, {Bournaud},
  {Bureau}, {Cappellari}, {Davies}, {Davis}, {de Zeeuw}, {Emsellem},
  {Khochfar}, {Krajnovi{\'c}}, {Kuntschner}, {Lablanche}, {McDermid},
  {Morganti}, {Naab}, {Oosterloo}, {Sarzi}, {Scott}, {Weijmans}, \&
  {Young}}]{Duc2011}
{Duc} P.-A. {et~al.}, 2011, \mnras, 417, 863

\bibitem[{{Emsellem} {et~al}\mbox{.}(2011){Emsellem}, {Cappellari},
  {Krajnovi{\'c}}, {Alatalo}, {Blitz}, {Bois}, {Bournaud}, {Bureau}, {Davies},
  {Davis}, {de Zeeuw}, {Khochfar}, {Kuntschner}, {Lablanche}, {McDermid},
  {Morganti}, {Naab}, {Oosterloo}, {Sarzi}, {Scott}, {Serra}, {van de Ven},
  {Weijmans}, \& {Young}}]{Emsellem2011}
{Emsellem} E. {et~al.}, 2011, \mnras, 414, 888

\bibitem[{{Gadotti}(2009)}]{Gadotti2009}
{Gadotti} D.~A., 2009, \mnras, 393, 1531

\bibitem[{{Graham} {et~al}\mbox{.}(1998){Graham}, {Colless}, {Busarello},
  {Zaggia}, \& {Longo}}]{Graham1998}
{Graham} A.~W., {Colless} M.~M., {Busarello} G., {Zaggia} S., {Longo} G., 1998,
  \aaps, 133, 325

\bibitem[{{Graham} \& {Driver}(2005)}]{Graham2005a}
{Graham} A.~W., {Driver} S.~P., 2005, \pasa, 22, 118

\bibitem[{{Graham} {et~al}\mbox{.}(2006){Graham}, {Merritt}, {Moore},
  {Diemand}, \& {Terzi{\'c}}}]{Graham2006b}
{Graham} A.~W., {Merritt} D., {Moore} B., {Diemand} J., {Terzi{\'c}} B., 2006,
  \aj, 132, 2711

\bibitem[{{Graham} \& {Worley}(2008)}]{Graham2008b}
{Graham} A.~W., {Worley} C.~C., 2008, \mnras, 388, 1708

\bibitem[{{Hill} {et~al}\mbox{.}(2011){Hill}, {Kelvin}, {Driver}, {Robotham},
  {Cameron}, {Cross}, {Andrae}, {Baldry}, {Bamford}, {Bland-Hawthorn},
  {Brough}, {Conselice}, {Dye}, {Hopkins}, {Liske}, {Loveday}, {Norberg},
  {Peacock}, {Croom}, {Frenk}, {Graham}, {Jones}, {Kuijken}, {Madore},
  {Nichol}, {Parkinson}, {Phillipps}, {Pimbblet}, {Popescu}, {Prescott},
  {Seibert}, {Sharp}, {Sutherland}, {Thomas}, {Tuffs}, \& {van
  Kampen}}]{Hill2011}
{Hill} D.~T. {et~al.}, 2011, \mnras, 412, 765

\bibitem[{{Kauffmann} {et~al}\mbox{.}(2003){Kauffmann}, {Heckman}, {White},
  {Charlot}, {Tremonti}, {Peng}, {Seibert}, {Brinkmann}, {Nichol}, {SubbaRao},
  \& {York}}]{Kauffmann2003b}
{Kauffmann} G. {et~al.}, 2003, \mnras, 341, 54

\bibitem[{{Kauffmann} {et~al}\mbox{.}(2004){Kauffmann}, {White}, {Heckman},
  {M{\'e}nard}, {Brinchmann}, {Charlot}, {Tremonti}, \&
  {Brinkmann}}]{Kauffmann2004}
{Kauffmann} G., {White} S.~D.~M., {Heckman} T.~M., {M{\'e}nard} B.,
  {Brinchmann} J., {Charlot} S., {Tremonti} C., {Brinkmann} J., 2004, \mnras,
  353, 713

\bibitem[{{Kelvin} {et~al}\mbox{.}(2010){Kelvin}, {Driver}, {Robotham}, {Hill},
  \& {Cameron}}]{Kelvin2010}
{Kelvin} L., {Driver} S., {Robotham} A., {Hill} D., {Cameron} E., 2010, in
  American Institute of Physics Conference Series, Vol. 1240, American
  Institute of Physics Conference Series, {Debattista} V.~P., {Popescu} C.~C.,
  eds., pp. 247--248

\bibitem[{{Kelvin} {et~al}\mbox{.}(2014){Kelvin}, {Driver}, {Robotham},
  {Graham}, {Phillipps}, {Agius}, {Alpaslan}, {Baldry}, {Bamford},
  {Bland-Hawthorn}, {Brough}, {Brown}, {Colless}, {Conselice}, {Hopkins},
  {Liske}, {Loveday}, {Norberg}, {Pimbblet}, {Popescu}, {Prescott}, {Taylor},
  \& {Tuffs}}]{Kelvin2014a}
{Kelvin} L.~S. {et~al.}, 2014, \mnras, 439, 1245

\bibitem[{{Kelvin} {et~al}\mbox{.}(2012){Kelvin}, {Driver}, {Robotham}, {Hill},
  {Alpaslan}, {Baldry}, {Bamford}, {Bland-Hawthorn}, {Brough}, {Graham},
  {H{\"a}ussler}, {Hopkins}, {Liske}, {Loveday}, {Norberg}, {Phillipps},
  {Popescu}, {Prescott}, {Taylor}, \& {Tuffs}}]{Kelvin2012}
{Kelvin} L.~S. {et~al.}, 2012, \mnras, 421, 1007

\bibitem[{{Kere{\v s}} {et~al}\mbox{.}(2005){Kere{\v s}}, {Katz}, {Weinberg},
  \& {Dav{\'e}}}]{Keres2005}
{Kere{\v s}} D., {Katz} N., {Weinberg} D.~H., {Dav{\'e}} R., 2005, \mnras, 363,
  2

\bibitem[{{Khochfar} {et~al}\mbox{.}(2011){Khochfar}, {Emsellem}, {Serra},
  {Bois}, {Alatalo}, {Bacon}, {Blitz}, {Bournaud}, {Bureau}, {Cappellari},
  {Davies}, {Davis}, {de Zeeuw}, {Duc}, {Krajnovi{\'c}}, {Kuntschner},
  {Lablanche}, {McDermid}, {Morganti}, {Naab}, {Oosterloo}, {Sarzi}, {Scott},
  {Weijmans}, \& {Young}}]{Khochfar2011}
{Khochfar} S. {et~al.}, 2011, \mnras, 417, 845

\bibitem[{{Khochfar} \& {Silk}(2006{\natexlab{a}})}]{Khochfar2006b}
{Khochfar} S., {Silk} J., 2006{\natexlab{a}}, \apjl, 648, L21

\bibitem[{{Khochfar} \& {Silk}(2006{\natexlab{b}})}]{Khochfar2006a}
{Khochfar} S., {Silk} J., 2006{\natexlab{b}}, \mnras, 370, 902

\bibitem[{{Kormendy} \& {Bender}(2012)}]{Kormendy2012}
{Kormendy} J., {Bender} R., 2012, \apjs, 198, 2

\bibitem[{{Krajnovi{\'c}} {et~al}\mbox{.}(2011){Krajnovi{\'c}}, {Emsellem},
  {Cappellari}, {Alatalo}, {Blitz}, {Bois}, {Bournaud}, {Bureau}, {Davies},
  {Davis}, {de Zeeuw}, {Khochfar}, {Kuntschner}, {Lablanche}, {McDermid},
  {Morganti}, {Naab}, {Oosterloo}, {Sarzi}, {Scott}, {Serra}, {Weijmans}, \&
  {Young}}]{Krajnovic2011}
{Krajnovi{\'c}} D. {et~al.}, 2011, \mnras, 414, 2923

\bibitem[{{Lara-L{\'o}pez} {et~al}\mbox{.}(2010){Lara-L{\'o}pez}, {Cepa},
  {Bongiovanni}, {P{\'e}rez Garc{\'{\i}}a}, {Ederoclite}, {Casta{\~n}eda},
  {Fern{\'a}ndez Lorenzo}, {Povi{\'c}}, \&
  {S{\'a}nchez-Portal}}]{Lara-Lopez2010}
{Lara-L{\'o}pez} M.~A. {et~al.}, 2010, \aap, 521, L53

\bibitem[{{Lara-L{\'o}pez} {et~al}\mbox{.}(2013){Lara-L{\'o}pez}, {Hopkins},
  {L{\'o}pez-S{\'a}nchez}, {Brough}, {Gunawardhana}, {Colless}, {Robotham},
  {Bauer}, {Bland-Hawthorn}, {Cluver}, {Driver}, {Foster}, {Kelvin}, {Liske},
  {Loveday}, {Owers}, {Ponman}, {Sharp}, {Steele}, {Taylor}, \&
  {Thomas}}]{Lara-Lopez2013}
{Lara-L{\'o}pez} M.~A. {et~al.}, 2013, \mnras, 434, 451

\bibitem[{{Lawrence} {et~al}\mbox{.}(2007){Lawrence}, {Warren}, {Almaini},
  {Edge}, {Hambly}, {Jameson}, {Lucas}, {Casali}, {Adamson}, {Dye}, {Emerson},
  {Foucaud}, {Hewett}, {Hirst}, {Hodgkin}, {Irwin}, {Lodieu}, {McMahon},
  {Simpson}, {Smail}, {Mortlock}, \& {Folger}}]{Lawrence2007}
{Lawrence} A. {et~al.}, 2007, \mnras, 379, 1599

\bibitem[{{L'Huillier} {et~al}\mbox{.}(2012){L'Huillier}, {Combes}, \&
  {Semelin}}]{LHuillier2012}
{L'Huillier} B., {Combes} F., {Semelin} B., 2012, \aap, 544, A68

\bibitem[{{Liske} {et~al}\mbox{.}(2003){Liske}, {Lemon}, {Driver}, {Cross}, \&
  {Couch}}]{Liske2003b}
{Liske} J., {Lemon} D.~J., {Driver} S.~P., {Cross} N.~J.~G., {Couch} W.~J.,
  2003, \mnras, 344, 307

\bibitem[{{Matkovi{\'c}} \& {Guzm{\'a}n}(2005)}]{Matkovic2005}
{Matkovi{\'c}} A., {Guzm{\'a}n} R., 2005, \mnras, 362, 289

\bibitem[{{Moustakas} {et~al}\mbox{.}(2013){Moustakas}, {Coil}, {Aird},
  {Blanton}, {Cool}, {Eisenstein}, {Mendez}, {Wong}, {Zhu}, \&
  {Arnouts}}]{Moustakas2013}
{Moustakas} J. {et~al.}, 2013, \apj, 767, 50

\bibitem[{{Muzzin} {et~al}\mbox{.}(2013){Muzzin}, {Marchesini}, {Stefanon},
  {Franx}, {McCracken}, {Milvang-Jensen}, {Dunlop}, {Fynbo}, {Brammer},
  {Labb{\'e}}, \& {van Dokkum}}]{Muzzin2013}
{Muzzin} A. {et~al.}, 2013, \apj, 777, 18

\bibitem[{{Navarro} \& {Benz}(1991)}]{Navarro1991}
{Navarro} J.~F., {Benz} W., 1991, \apj, 380, 320

\bibitem[{{Omand} {et~al}\mbox{.}(2014){Omand}, {Balogh}, \&
  {Poggianti}}]{Omand2014}
{Omand} C., {Balogh} M., {Poggianti} B., 2014, arXiv:1402.3394

\bibitem[{{Pastrav} {et~al}\mbox{.}(2013){Pastrav}, {Popescu}, {Tuffs}, \&
  {Sansom}}]{Pastrav2013a}
{Pastrav} B.~A., {Popescu} C.~C., {Tuffs} R.~J., {Sansom} A.~E., 2013, \aap,
  553, A80

\bibitem[{{Patel} {et~al}\mbox{.}(2013){Patel}, {van Dokkum}, {Franx},
  {Quadri}, {Muzzin}, {Marchesini}, {Williams}, {Holden}, \&
  {Stefanon}}]{Patel2013}
{Patel} S.~G. {et~al.}, 2013, \apj, 766, 15

\bibitem[{{Peng} {et~al}\mbox{.}(2002){Peng}, {Ho}, {Impey}, \&
  {Rix}}]{Peng2002}
{Peng} C.~Y., {Ho} L.~C., {Impey} C.~D., {Rix} H.-W., 2002, \aj, 124, 266

\bibitem[{{Peng} {et~al}\mbox{.}(2010{\natexlab{a}}){Peng}, {Ho}, {Impey}, \&
  {Rix}}]{Peng2010a}
{Peng} C.~Y., {Ho} L.~C., {Impey} C.~D., {Rix} H.-W., 2010{\natexlab{a}}, \aj,
  139, 2097

\bibitem[{{Peng} {et~al}\mbox{.}(2010{\natexlab{b}}){Peng}, {Lilly}, {Kova{\v
  c}}, {Bolzonella}, {Pozzetti}, {Renzini}, {Zamorani}, {Ilbert}, {Knobel},
  {Iovino}, {Maier}, {Cucciati}, {Tasca}, {Carollo}, {Silverman}, {Kampczyk},
  {de Ravel}, {Sanders}, {Scoville}, {Contini}, {Mainieri}, {Scodeggio},
  {Kneib}, {Le F{\`e}vre}, {Bardelli}, {Bongiorno}, {Caputi}, {Coppa}, {de la
  Torre}, {Franzetti}, {Garilli}, {Lamareille}, {Le Borgne}, {Le Brun},
  {Mignoli}, {Perez Montero}, {Pello}, {Ricciardelli}, {Tanaka}, {Tresse},
  {Vergani}, {Welikala}, {Zucca}, {Oesch}, {Abbas}, {Barnes}, {Bordoloi},
  {Bottini}, {Cappi}, {Cassata}, {Cimatti}, {Fumana}, {Hasinger}, {Koekemoer},
  {Leauthaud}, {Maccagni}, {Marinoni}, {McCracken}, {Memeo}, {Meneux}, {Nair},
  {Porciani}, {Presotto}, \& {Scaramella}}]{Peng2010b}
{Peng} Y.-j. {et~al.}, 2010{\natexlab{b}}, \apj, 721, 193

\bibitem[{{Peng} {et~al}\mbox{.}(2012){Peng}, {Lilly}, {Renzini}, \&
  {Carollo}}]{Peng2012}
{Peng} Y.-j., {Lilly} S.~J., {Renzini} A., {Carollo} M., 2012, \apj, 757, 4

\bibitem[{{Pichon} {et~al}\mbox{.}(2011){Pichon}, {Pogosyan}, {Kimm}, {Slyz},
  {Devriendt}, \& {Dubois}}]{Pichon2011}
{Pichon} C., {Pogosyan} D., {Kimm} T., {Slyz} A., {Devriendt} J., {Dubois} Y.,
  2011, \mnras, 418, 2493

\bibitem[{{Pozzetti} {et~al}\mbox{.}(2010){Pozzetti}, {Bolzonella}, {Zucca},
  {Zamorani}, {Lilly}, {Renzini}, {Moresco}, {Mignoli}, {Cassata}, {Tasca},
  {Lamareille}, {Maier}, {Meneux}, {Halliday}, {Oesch}, {Vergani}, {Caputi},
  {Kova{\v c}}, {Cimatti}, {Cucciati}, {Iovino}, {Peng}, {Carollo}, {Contini},
  {Kneib}, {Le F{\'e}vre}, {Mainieri}, {Scodeggio}, {Bardelli}, {Bongiorno},
  {Coppa}, {de la Torre}, {de Ravel}, {Franzetti}, {Garilli}, {Kampczyk},
  {Knobel}, {Le Borgne}, {Le Brun}, {Pell{\`o}}, {Perez Montero},
  {Ricciardelli}, {Silverman}, {Tanaka}, {Tresse}, {Abbas}, {Bottini}, {Cappi},
  {Guzzo}, {Koekemoer}, {Leauthaud}, {Maccagni}, {Marinoni}, {McCracken},
  {Memeo}, {Porciani}, {Scaramella}, {Scarlata}, \& {Scoville}}]{Pozzetti2010}
{Pozzetti} L. {et~al.}, 2010, \aap, 523, A13

\bibitem[{{Robotham} {et~al}\mbox{.}(2013){Robotham}, {Liske}, {Driver},
  {Sansom}, {Baldry}, {Bauer}, {Bland-Hawthorn}, {Brough}, {Brown}, {Colless},
  {Christodoulou}, {Drinkwater}, {Grootes}, {Hopkins}, {Kelvin}, {Norberg},
  {Loveday}, {Phillipps}, {Sharp}, {Taylor}, \& {Tuffs}}]{Robotham2013}
{Robotham} A.~S.~G. {et~al.}, 2013, \mnras, 431, 167

\bibitem[{{Robotham} {et~al}\mbox{.}(2011){Robotham}, {Norberg}, {Driver},
  {Baldry}, {Bamford}, {Hopkins}, {Liske}, {Loveday}, {Merson}, {Peacock},
  {Brough}, {Cameron}, {Conselice}, {Croom}, {Frenk}, {Gunawardhana}, {Hill},
  {Jones}, {Kelvin}, {Kuijken}, {Nichol}, {Parkinson}, {Pimbblet}, {Phillipps},
  {Popescu}, {Prescott}, {Sharp}, {Sutherland}, {Taylor}, {Thomas}, {Tuffs},
  {van Kampen}, \& {Wijesinghe}}]{Robotham2011b}
{Robotham} A.~S.~G. {et~al.}, 2011, \mnras, 416, 2640

\bibitem[{{Schechter}(1976)}]{Schechter1976}
{Schechter} P., 1976, \apj, 203, 297

\bibitem[{{Schmidt}(1968)}]{Schmidt1968}
{Schmidt} M., 1968, \apj, 151, 393

\bibitem[{{S{\'e}rsic}(1963)}]{Sersic1963}
{S{\'e}rsic} J.~L., 1963, Boletin de la Asociacion Argentina de Astronomia La
  Plata Argentina, 6, 41

\bibitem[{{Shankar} {et~al}\mbox{.}(2013){Shankar}, {Marulli}, {Bernardi},
  {Mei}, {Meert}, \& {Vikram}}]{Shankar2013}
{Shankar} F., {Marulli} F., {Bernardi} M., {Mei} S., {Meert} A., {Vikram} V.,
  2013, \mnras, 428, 109

\bibitem[{{Shen} {et~al}\mbox{.}(2003){Shen}, {Mo}, {White}, {Blanton},
  {Kauffmann}, {Voges}, {Brinkmann}, \& {Csabai}}]{Shen2003}
{Shen} S., {Mo} H.~J., {White} S.~D.~M., {Blanton} M.~R., {Kauffmann} G.,
  {Voges} W., {Brinkmann} J., {Csabai} I., 2003, \mnras, 343, 978

\bibitem[{{Shimizu} \& {Inoue}(2013)}]{Shimizu2013}
{Shimizu} I., {Inoue} A.~K., 2013, arXiv:1310.0879

\bibitem[{{Simard} {et~al}\mbox{.}(2011){Simard}, {Mendel}, {Patton},
  {Ellison}, \& {McConnachie}}]{Simard2011}
{Simard} L., {Mendel} J.~T., {Patton} D.~R., {Ellison} S.~L., {McConnachie}
  A.~W., 2011, \apjs, 196, 11

\bibitem[{{Szomoru} {et~al}\mbox{.}(2013){Szomoru}, {Franx}, {van Dokkum},
  {Trenti}, {Illingworth}, {Labb{\'e}}, \& {Oesch}}]{Szomoru2013}
{Szomoru} D., {Franx} M., {van Dokkum} P.~G., {Trenti} M., {Illingworth} G.~D.,
  {Labb{\'e}} I., {Oesch} P., 2013, \apj, 763, 73

\bibitem[{{Tasca} \& {White}(2011)}]{Tasca2011}
{Tasca} L.~A.~M., {White} S.~D.~M., 2011, \aap, 530, A106

\bibitem[{{Taylor} {et~al}\mbox{.}(2011){Taylor}, {Hopkins}, {Baldry}, {Brown},
  {Driver}, {Kelvin}, {Hill}, {Robotham}, {Bland-Hawthorn}, {Jones}, {Sharp},
  {Thomas}, {Liske}, {Loveday}, {Norberg}, {Peacock}, {Bamford}, {Brough},
  {Colless}, {Cameron}, {Conselice}, {Croom}, {Frenk}, {Gunawardhana},
  {Kuijken}, {Nichol}, {Parkinson}, {Phillipps}, {Pimbblet}, {Popescu},
  {Prescott}, {Sutherland}, {Tuffs}, {van Kampen}, \&
  {Wijesinghe}}]{Taylor2011}
{Taylor} E.~N. {et~al.}, 2011, \mnras, 418, 1587

\bibitem[{{Tomczak} {et~al}\mbox{.}(2014){Tomczak}, {Quadri}, {Tran},
  {Labb{\'e}}, {Straatman}, {Papovich}, {Glazebrook}, {Allen}, {Brammer},
  {Kacprzak}, {Kawinwanichakij}, {Kelson}, {McCarthy}, {Mehrtens}, {Monson},
  {Persson}, {Spitler}, {Tilvi}, \& {van Dokkum}}]{Tomczak2014}
{Tomczak} A.~R. {et~al.}, 2014, \apj, 783, 85

\bibitem[{{Tremonti} {et~al}\mbox{.}(2004){Tremonti}, {Heckman}, {Kauffmann},
  {Brinchmann}, {Charlot}, {White}, {Seibert}, {Peng}, {Schlegel}, {Uomoto},
  {Fukugita}, \& {Brinkmann}}]{Tremonti2004}
{Tremonti} C.~A. {et~al.}, 2004, \apj, 613, 898

\bibitem[{{van den Bosch} {et~al}\mbox{.}(2008){van den Bosch}, {Aquino},
  {Yang}, {Mo}, {Pasquali}, {McIntosh}, {Weinmann}, \&
  {Kang}}]{vandenBosch2008}
{van den Bosch} F.~C., {Aquino} D., {Yang} X., {Mo} H.~J., {Pasquali} A.,
  {McIntosh} D.~H., {Weinmann} S.~M., {Kang} X., 2008, \mnras, 387, 79

\bibitem[{{van Dokkum} {et~al}\mbox{.}(2010){van Dokkum}, {Whitaker},
  {Brammer}, {Franx}, {Kriek}, {Labb{\'e}}, {Marchesini}, {Quadri}, {Bezanson},
  {Illingworth}, {Muzzin}, {Rudnick}, {Tal}, \& {Wake}}]{vanDokkum2010a}
{van Dokkum} P.~G. {et~al.}, 2010, \apj, 709, 1018

\bibitem[{{White} \& {Frenk}(1991)}]{White1991}
{White} S.~D.~M., {Frenk} C.~S., 1991, \apj, 379, 52

\bibitem[{{Wyse} {et~al}\mbox{.}(1997){Wyse}, {Gilmore}, \& {Franx}}]{Wyse1997}
{Wyse} R.~F.~G., {Gilmore} G., {Franx} M., 1997, \araa, 35, 637

\bibitem[{{Yang} {et~al}\mbox{.}(2009){Yang}, {Mo}, \& {van den
  Bosch}}]{Yang2009}
{Yang} X., {Mo} H.~J., {van den Bosch} F.~C., 2009, \apj, 695, 900

\bibitem[{{York} {et~al}\mbox{.}(2000){York}, {Adelman}, {Anderson},
  {Anderson}, {Annis}, {Bahcall}, {Bakken}, {Barkhouser}, {Bastian}, {Berman},
  {Boroski}, {Bracker}, {Briegel}, {Briggs}, {Brinkmann}, {Brunner}, {Burles},
  {Carey}, {Carr}, {Castander}, {Chen}, {Colestock}, {Connolly}, {Crocker},
  {Csabai}, {Czarapata}, {Davis}, {Doi}, {Dombeck}, {Eisenstein}, {Ellman},
  {Elms}, {Evans}, {Fan}, {Federwitz}, {Fiscelli}, {Friedman}, {Frieman},
  {Fukugita}, {Gillespie}, {Gunn}, {Gurbani}, {de Haas}, {Haldeman}, {Harris},
  {Hayes}, {Heckman}, {Hennessy}, {Hindsley}, {Holm}, {Holmgren}, {Huang},
  {Hull}, {Husby}, {Ichikawa}, {Ichikawa}, {Ivezi{\'c}}, {Kent}, {Kim},
  {Kinney}, {Klaene}, {Kleinman}, {Kleinman}, {Knapp}, {Korienek}, {Kron},
  {Kunszt}, {Lamb}, {Lee}, {Leger}, {Limmongkol}, {Lindenmeyer}, {Long},
  {Loomis}, {Loveday}, {Lucinio}, {Lupton}, {MacKinnon}, {Mannery}, {Mantsch},
  {Margon}, {McGehee}, {McKay}, {Meiksin}, {Merelli}, {Monet}, {Munn},
  {Narayanan}, {Nash}, {Neilsen}, {Neswold}, {Newberg}, {Nichol}, {Nicinski},
  {Nonino}, {Okada}, {Okamura}, {Ostriker}, {Owen}, {Pauls}, {Peoples},
  {Peterson}, {Petravick}, {Pier}, {Pope}, {Pordes}, {Prosapio},
  {Rechenmacher}, {Quinn}, {Richards}, {Richmond}, {Rivetta}, {Rockosi},
  {Ruthmansdorfer}, {Sandford}, {Schlegel}, {Schneider}, {Sekiguchi}, {Sergey},
  {Shimasaku}, {Siegmund}, {Smee}, {Smith}, {Snedden}, {Stone}, {Stoughton},
  {Strauss}, {Stubbs}, {SubbaRao}, {Szalay}, {Szapudi}, {Szokoly}, {Thakar},
  {Tremonti}, {Tucker}, {Uomoto}, {Vanden Berk}, {Vogeley}, {Waddell}, {Wang},
  {Watanabe}, {Weinberg}, {Yanny}, {Yasuda}, \& {SDSS
  Collaboration}}]{York2000}
{York} D.~G. {et~al.}, 2000, \aj, 120, 1579

\bibitem[{{Young} \& {Currie}(1994)}]{Young1994}
{Young} C.~K., {Currie} M.~J., 1994, \mnras, 268, L11

\end{thebibliography}

\appendix

\section{Impact of the LBS Population on the GSMF}

\label{app:lbs}
\begin{figure*}
\includegraphics[width=1\textwidth]{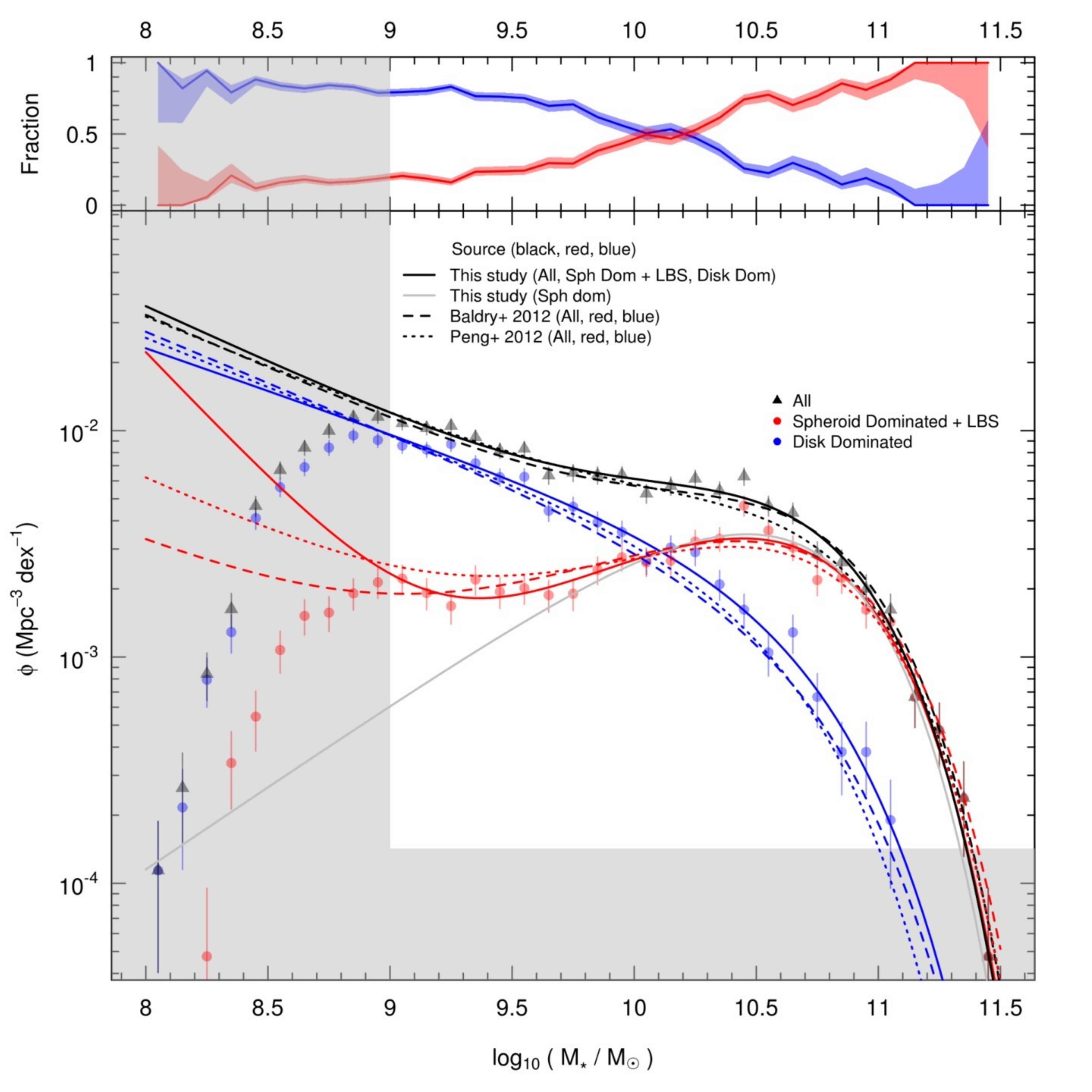}

\caption{\label{fig:massfunctypeLBS}As Figure \ref{fig:massfunctype}, but
with the inclusion of the LBS population in the spheroid dominated
class (red data points). The combined spheroid dominated plus LBS
population is fitted by a double Schechter component (red solid line)
as is the total galaxy population (solid black line), whereas the
disk dominated population remains well described by a single Schechter
function (solid blue line). The previous spheroid dominated (E, S0-Sa)
single Schechter function fit is shown in light grey, for reference.
Comparison Schechter function fits for similar red and blue populations
from \citet{Baldry2012} and \citet{Peng2012} are also shown.}
\end{figure*}

Our prior division of our galaxy sample into spheroid
dominated (E, S0-Sa) and disk dominated (Sab-Scd, Sd-Irr) galaxies,
as shown in Figure \ref{fig:massfunctype}, neglected the low-mass
little blue spheroid population. Figure \ref{fig:massfunctypeLBS}
shows the GSMF and disk dominated MSMF as before, but with an updated
spheroid dominated MSMF including the LBS population (i.e., E, S0-Sa,
LBS). All data analysis is conducted in a similar fashion to that
outlined in Section \ref{sec:massfuncs}. The previous spheroid dominated
(E, S0-Sa) single Schechter function fit is shown in light grey, for
reference. As can clearly be seen, once the LBS galaxy population
is included into the spheroid dominated class, we recover a low-mass
upturn exceedingly similar in nature to the red population as reported
in, e.g., \citet{Baldry2012} and \citet{Peng2012}. On the surface,
the spheroid dominated class may perhaps be the natural home of the
`little blue spheroid' galaxy population, allowing us to maintain
a good level of agreement with comparison studies. 

However, we remind the reader that our adopted visual morphological
classification boundaries \citep{Kelvin2014a} are substantially different
from the red/blue divisions presented in \citet{Baldry2012} and \citet{Peng2012},
and also the star forming/quiescent divisions as noted in Section
\ref{sub:msmf}. See, for example, Figure \ref{fig:masscolourgrid}
for a visual representation of the colour mix across all morphologies.
Indeed, despite the inherent trends between morphology, colour and
star formation rate, we see no explicit reason why a bimodal division
along morphological lines should reproduce exactly that of one which
has been created along colour or star formation rate measures. In
which case, it is perhaps surprising that a combined spheroid dominated
plus LBS population so closely recovers the low-mass upturn observed
in the red populations of \citet{Baldry2012} and \citet{Peng2012}.
Also note that whilst the third word in LBS denotes its shape, the
second part of the acronym denotes their typical colour: blue. As
is shown in Figure \ref{fig:masscolourgrid}, the majority of blue
galaxies lie in the disk dominated Sab-Scd and Sd-Irr classes, giving
weight to the inclusion of the LBS population in our disk dominated
sub-sample instead. This would only serve to increase the low-mass
upturn of the disk dominated population, and maintain the low-mass
discrepancy we observe between our spheroid dominated class and the
comparison red-population data from the literature. The correct placement
of our LBS galaxy population within the morphological schema adopted
throughout this study remains unclear, and therefore, we continue
to advocate its exclusion at present. Future studies are planned to
clarify the importance of the LBS population (Moffett et al., in prep.).

Table \ref{tab:massfuncmorphLBS} provides the double Schechter fit
parameters to the combined spheroid dominated (E, S0-Sa) plus LBS
population. Note the unusually low $\alpha_{2}$ slope parameter,
combined with relatively large error bars. This indicates that shape
of the low-mass end of our Schechter fit is poorly constrained, as
is evidenced by the unusually steep gradient of the fit when extrapolated
below our mass limit (see Figure \ref{fig:massfunctypeLBS}). Nevertheless,
the Schechter fit provides a good description of the data across the
range of interest ($\log\left(\mathcal{M}_{*}/\mathcal{M}_{\odot}\right)>9.0$),
exhibiting a strong goodness of fit parameter.

\renewcommand{\arraystretch}{1.5}
\setlength{\tabcolsep}{6pt}

\begin{table*}
\begin{centering}
\begin{tabular}{cccccccc}
\hline 
$\log\left(\mathcal{M}^{*}/\mathcal{M}_{\odot}\right)$ & $\alpha_{1}$ & $\phi_{1}^{*}/10^{-3}$ & $\alpha_{2}$ & $\phi_{2}^{*}/10^{-3}$ & $\chi^{2}/\nu$ & $\delta_{\phi}/10^{7}$ & $\delta_{\Sigma}/10^{7}$\tabularnewline
 &  & ($\mathrm{dex}^{-1}\mathrm{Mpc}^{-3}$) &  & ($\mathrm{dex}^{-1}\mathrm{Mpc}^{-3}$) &  & ($\mathcal{M}_{\odot}\mathrm{Mpc^{-3}}$) & ($\mathcal{M}_{\odot}\mathrm{Mpc^{-3}}$) \tabularnewline
\hline 
$10.65\pm0.08$ & $-0.37\pm0.23$ & $3.63\pm1.38$ & $-2.13\pm1.23$ & $0.01\pm0.05$ & $0.96$ & $14.29\pm7.76$ & $15.56_{-3.74}^{+4.95}$ \tabularnewline
\hline 
\end{tabular}
\par\end{centering}

\caption{\label{tab:massfuncmorphLBS}Double Schechter stellar mass function
fit parameters for the combined spheroid dominated plus LBS galaxy
population as shown in Figure \ref{fig:massfunctypeLBS}. From left
to right, columns are: the shared knee in the Schechter function ($\mathcal{M}^{*}$);
the primary slope of the faint end of the Schechter function ($\alpha_{1}$);
the primary normalisation constant for the Schechter function fit
($\phi_{1}^{*}$); the secondary slope of the faint end of the Schechter
function ($\alpha_{2}$); the secondary normalisation constant for
the Schechter function fit ($\phi_{2}^{*}$); the $\chi^{2}$ goodness
of fit parameter ($\chi^{2}/\nu$); the stellar mass density implied
in the usual way via the fitted Schechter function {[}$\rho_{\phi}=\sum_{i=1}^{N}\phi_{i}^{*}\mathcal{M}^{*}\Gamma\left(\alpha_{i}+2\right)${]};
the stellar mass density calculated via the direct summation of the
stellar masses from the individual galaxies {[}$\rho_{\Sigma}=\frac{1}{V}\sum_{i=1}^{N}\mathcal{M}_{*}${]}.
The double Schechter function is fit to $N=24$ data bins with $n=5$
fitted parameters, therefore, the number of degrees of freedom for
this fit is given by $\nu=N-n=19$.}
\end{table*}

\renewcommand{\arraystretch}{1}

\label{lastpage}
\end{document}